\documentclass[prc,aps,floatfix,twocolumn]{revtex4-1}
\usepackage{dcolumn}
\usepackage{slashed}
\usepackage{bm}
\usepackage{graphicx}
\usepackage[utf8]{inputenc} 							% lettere accentate da tastiera
\usepackage{bbm}
\usepackage{simplewick}
\usepackage{xcolor}
\usepackage[normalem]{ulem}
\usepackage{multirow}%
\usepackage{amsmath,amssymb,amsfonts}%
\usepackage{amsthm}%
\usepackage{mathrsfs}%
\usepackage[title]{appendix}%
\usepackage{textcomp}%
\usepackage{booktabs}%
\usepackage{afterpage}
\usepackage{listings}%
\usepackage{bbold}
\usepackage{braket}
\def\grad{\nabla}
\def\dag{\dagger}

\def\beo{{{}^8{\rm Be}}}
\def\C{{{}^{12}{\rm C}}}
\def\Ox{{{}^{16}{\rm O}}}
\def\a{\alpha}

\def\bmx{{\bm x}}

\def\bmp{{\bm p}}

\def\bmK{{\bm K}}

\def\bmQ{{\bm Q}}

\begin{document}
%%%%%%%%%%%%%%%%%%%%%%%%%%%%%%%%%%%%%%%%%%%%%%%%%%%%%%%%%%%%%%%%%%%%%%%%%%%%%%%%%%%%
\title{The 3-$\a$ and 4-$\a$ particle systems within short-range Effective Field Theory}
%%%%%%%%%%%%%%%%%%%%%%%%%%%%%%%%%%%%%%%%%%%%%%%%%%%%%%%%%%%%%%%%%%%%%%%%%%%%%%%%%%%%%%%%%%%%%%
\author{ E. Filandri$^{1,2}$,  M. Viviani$^2$,  L. Girlanda$^{3,4}$, A. Kievsky$^2$ and
L.E. Marcucci$^{1,2}$}

\affiliation{
$^1$Department of Physics ``E. Fermi'', University of Pisa, I-56127 Pisa, Italy\\
$^2$INFN-Pisa, I-56127, Pisa, Italy \\
$^3$Department of Mathematics and Physics ``E. De Giorgi", University of Salento, I-73100 Lecce, Italy \\
$^4$INFN-Lecce, I-73100 Lecce, Italy  }

\date{\today}
\begin{abstract}
$\C$ and $\Ox$  nuclei represent essential elements for life on Earth. The study of their formation plays a key role in understanding heavy element nucleosynthesis and stellar evolution. In this paper we present the study of $\C$ and $\Ox$ nuclei as systems composed of $\a$-particle clusters using the  short-range effective field theory  approach.
The fundamental and excited states of the studied nuclei are calculated within an  ab-initio approach, using the Hyperspherical Harmonics method.
Thanks to the two-body potential and fine-tuning of the three-body force, we have found  the $\C$ system nicely reproduced by theory. However, for the $\Ox$ case, it is necessary to include a 4-body force in order to achieve agreement with the experimental data.
\end{abstract}


\maketitle
\section{Introduction}
The nucleosynthesis of Carbon ($\C$) and Oxygen ($\Ox$) in the universe stands as a fundamental issue within nuclear astrophysics, owing to its implications for understanding the origins of life on Earth and the life cycles of stars. Consequently, an accurate characterization of these systems serves as a crucial starting point for subsequent analyses of the  reactions that drive their production.

The theoretical concept that these nuclei may consist of $\alpha$-clusters  has a rich history, dating back to Rutherford's discovery of $\a$ radiation and the subsequent development of quantum mechanics (see Ref.~\cite{Freer} and references therein). Despite this early inception, recent advances in radioactive ion beam experiments have opened a new avenue of research, wherein experimental data are compared to theoretical predictions with or without cluster assumptions. Cluster structures arise from a delicate balance among repulsive short-range forces, Pauli blocking effects, attractive medium-range nuclear interactions, and long-range Coulomb potential. In this equilibrium, four nucleons tend to aggregate, forming the spin-zero entity known as the nucleus of $^4$He, commonly referred to as the $\a$-particle. Notably, among light nuclei, the $\alpha$-particle exhibits the highest binding energy per nucleon (approximately $7$ MeV), suggesting that it is a natural state for nucleon aggregation.

The energy required to dissociate these systems is considerably lower than the excitation energy of the $\alpha$ particle, indicating a separation of energy scales. Therefore, the requisite condition for the development of an effective field theory (EFT) framework is satisfied.
Effective field theories capture the most general interactions among low-energy degrees of freedom consistent with certain assumed symmetries. The Cluster EFT~\cite{Bert2002,HIGA08,Chenhalo,Bed2003} represents a variant of the well-established pionless EFT~\cite{van_Kolck_1998,van_Kolck_1999,Kaplan_1998} and is suitable for describing bound states and reactions involving halo or cluster nuclei, in an energy range where clusters of nucleons can be treated as elementary degrees of freedom. While certain intricate aspects of nuclear structure remain beyond the scope of this approach, it facilitates the study of ground state properties of cluster nuclei and certain low-energy astrophysical reactions involving them.

The application of the cluster approach to investigate $\C$ and $\Ox$ is not new. Previous studies have characterized these nuclei as 3-$\alpha$ and 4-$\alpha$ systems, employing phenomenological potentials for ground state calculations (see as an example Refs. \cite{1974PThPh..52..890S,Ronen:2005xs}). In this study, on the other hand, we derive interactions derived from Cluster EFT, therefore grounding them in a more robust theoretical framework. Then, we use these interactions to study the ground- and excited-state energy levels solving the Schr$\mathrm{\ddot{o}}$dinger equation, through the variational principle, from the diagonalization of the Hamiltonian expanded on the Hyperspherical Harmonics (HH) basis. 

Having sketched our work, we organize the paper as follows: in Sec. \ref{sec:theory} the theoretical background of this  work is presented. In  particular we describe in Sec. \ref{sec:theorya} the interactions arising from the Cluster EFT and in Sec. \ref{sec:theoryb} the HH method used to obtain the various energy levels.
In Sec. \ref{sec:results} we present our results for the $\C$, $\Ox$ ground and excited states. In Sec. \ref{sec:conclusion} we draw conclusions and provide an outlook for this work.

\section{Theoretical background}\label{sec:theory}
\subsection{The $\a$-Cluster EFT approach}\label{sec:theorya}

In order to construct the interaction terms of the $\alpha$-$\alpha$ EFT Hamiltonian, we adopt the so-called 'bottom-up' approach, considering  all the constraints imposed by the symmetries of the underlying theory. For the low-energy contact theory, we focus on the non-relativistic Quantum Chromodynamic (QCD) spatial symmetries, including parity, time-reversal, permutational symmetries, and Galilean invariance. Since we are operating in a non-relativistic regime, we can ignore antiparticles.
Therefore the  scalar fields representing $\alpha$ particles are described  in momentum space by
\begin{align}
    \psi(x)&=\int \frac{d\bm{q}}{(2\pi)^3}a_{\bm{q}}e^{-i\bm{q}\bmx}\,,\label{eq:field1}\\
       \psi(x)^{\dag}&=\int \frac{d\bm{q}}{(2\pi)^3}a^{\dag}_{\bm{q}}e^{i\bm{q}\bmx}\,,\label{eq:field}
\end{align}
with $a_{\bm{q}}$ ($a^{\dag}_{\bm{q}}$) being the creation (annihilation) particle operators.
We justify the use of non-relativistic theory as long as the center-of-mass energy ($E_{CM}$) is much smaller than the $\alpha$ particle mass ($m$). Excited states of $\alpha$ particles are not considered in this EFT, as their energy levels are significantly higher, typically around 20 MeV above the ground state. Therefore, the constraints on $E_{CM}$ and   the relative momentum  are crucial.

Up to  next-to-next-to-leading order (N2LO) the interaction terms involve even powers of gradient operators up to the fourth power, satisfying normal ordering and symmetry constraints. The resulting interaction Hamiltonian density is given by \cite{recchia},
\begin{align}
H_{\text {eff }}^{\text {strong }}=&\tilde{C_1}\psi^{\dag}\psi\psi^{\dag}\psi+\tilde{C_2}\grad(\psi^{\dag}\psi)\grad(\psi^{\dag}\psi)\nonumber\\&+\tilde{C_3}(\partial_i\partial_j(\psi^{\dag}\psi)(\psi^{\dag}\overleftrightarrow{\partial}_i\overleftrightarrow{\partial}_j\psi)\nonumber\\ &\qquad-\partial_i\partial_j(\psi^{\dag}\overleftrightarrow{\partial}_i\overleftrightarrow{\partial}_j\psi)(\psi^{\dag}\psi))\nonumber\\
&+\tilde{C_4}\grad^4(\psi^{\dag}\psi)\psi^{\dag}\psi\nonumber\\&+\tilde{C_5}(\grad^2(\psi^{\dag}\psi)\psi^{\dag}\overleftrightarrow{\grad}^2\psi\nonumber\\ &\qquad-\grad^2(\psi^{\dag}\overleftrightarrow{\grad}^2\psi)\psi^{\dag}\psi)\,,
\label{eq:Heff}
\end{align}
where $\tilde{C}_{i=1,5}$ are the low energy constants (LECs), having dimensions of some inverse powers of the energy. In order to make these parameters dimensionless, we have to multiply them by a proper scale factor $\Lambda$. 
From the effective Hamiltonian, we derive the momentum-space potential as
\begin{align}
V_{\text {eff }}&^{\text {strong }}(\boldsymbol{K}, \boldsymbol{Q})=C'_1+C'_2 \boldsymbol{K}^2\nonumber\\+&C'_3(\boldsymbol{K} \times \boldsymbol{Q})^2+C'_4 \boldsymbol{K}^4+C'_5 \boldsymbol{K}^2 \boldsymbol{Q}^2\,,\label{eq:V}
\end{align}
where $\bmK=\bmp'-\bmp$, $\bmQ=(\bmp+\bmp')/2$, $\bmp$ and $\bmp'$ being the initial and final  relative momenta. In Eq. (\ref{eq:V}) the  LECs $C'_{i=1,5}$ are combinations of the ones appearing in Eq. (\ref{eq:Heff}).

To address potential divergences, we employ a Gaussian regulator of the form $f_{\Lambda}\left(\bm{K}^2\right)=\mathrm{e}^{-\frac{K^2}{2 \Lambda^2}}\,$.
We then  perform the Fourier transform of the smeared strong interaction potential, in order to move to coordinate space obtaining 
\begin{align}
 V_{\text {eff }}^{\text {strong }}(\boldsymbol{r})&= C_1 \delta_a^{(3)}(\boldsymbol{r})+a^2 C_2 \nabla^2 \delta_a^{(3)}(\boldsymbol{r}) \nonumber\\
&+a^4[C_3\left(\frac{L(L+1)}{a^4}+\frac{2}{a^2}\left(\frac{1}{2} \stackrel{\leftrightarrow}{\nabla}\right)^2\delta_a^{(3)}(\boldsymbol{r})\right)\nonumber\\&+C_4 \nabla^4 \delta_a^{(3)}(\boldsymbol{r})-C_5 \nabla^2 \delta_a^{(3)}(\boldsymbol{r})\left(\frac{1}{2} \overleftrightarrow{\nabla}\right)^2]\,,\label{eq:Veff}
\end{align}
where  $\delta_a^{(3)}(\boldsymbol{r})=e^{-(r^2/2a^2)}$ , $L$ indicates the orbital angular quantum number and $a=\hbar c/\Lambda$. The LECs in Eq. (\ref{eq:Veff}) absorb the appropriate powers of the cutoff and are given in MeV unit.

In addition to the strong interaction, $\alpha$ particles interact via Coulomb repulsion. To cure the singularity at the origin of the $1/r$ Coulomb potential and to take  into account the $\a$ particle structure, we multiply the Coulomb momentum space potential by a  Gaussian regulator determined by $r_\a=1.44$ fm  as in Ref. \cite{ALI196699}. Then, we perform the Fourier transform.
The total $\alpha$-$\alpha$ potential  is obtained by adding the smeared Coulomb potential to the strong interaction potential, resulting in
\begin{align}
 V_{\text {eff }}(\boldsymbol{r})&=\frac{4 \alpha}{r} \operatorname{erf}\left(\frac{\sqrt{3}r}{2r_\a }\right)+  V_{\text {eff }}^{\text {strong }}(\boldsymbol{r})\,,\label{eq:Vtot}
\end{align}
where $\operatorname{erf}(x)$ is the error function as defined in Ref. \cite{abramowitz+stegun}.
In this study, for simplicity, we drop the non-local terms.

As required by EFT power counting, we also include  a three-body force, that we choose of the form
\begin{equation}
    V_3(\rho)=(V_{03}\hat{P}_{L=0}+V_{23}\hat{P}_{L=2})e^{-(\rho^2/2a_3^2)}\label{3bforce}\,,
\end{equation}
where $\hat{P}_{L=0,2}$ are the $L=0,2$ waves projectors and $\rho$ is the $A=3$ hyperradius defined as in Eq. (\ref{eq:rho}) below. We then tune the $V_{03},V_{23}$ and $a_3$  parameters of Eq. (\ref{3bforce}) on the  binding energy of the $\C$ ground state,   $2^+$ excited state and Hoyle state, respectively.

Furthermore, in order  to bridge  the gap in energies between theoretical predictions and experimental data for $\Ox$, we include in our model  a  four-body force of the form
\begin{equation}
    V_4(\rho)=(V_{04}\hat{P}_{L=0}+V_{24}\hat{P}_{L=2})e^{-(\rho^2/2a_4^2)}\label{4bforce}\,
\end{equation}
where $L$ and $\rho$ are the four-body total angular momentum and hyperradius, respectively, $\hat{P}_{L=0,2}$ is defined as before and the parameters $V_{04},V_{24}$ and $a_4$  are fitted to the binding energy of the  $\Ox$ ground state, first  and  $2^+$ excited state.

\subsection{The HH method}\label{sec:theoryb}
\subsubsection{The HH functions}

Let us consider a system of $A$ bosons with masses $m_{i=1,\cdots,A}$ and spatial coordinates $\boldsymbol{r}_1, \ldots, \boldsymbol{r}_A$, respectively. To separate the internal and center-of-mass (CM) motion, it is convenient to introduce another set of coordinates consisting of $N=A-1$ internal Jacobi coordinates $\bmx_1, \ldots, \bmx_N$ and the CM coordinate $\boldsymbol{X}$ defined by
\begin{equation}
\boldsymbol{X}=\frac{1}{M} \sum_{i=1}^A m_i \boldsymbol{r}_i,
\end{equation}
where $M=\sum_{i=1}^A m_i$ is the total mass of the system. 
From now on we consider all masses to be equal, i.e. $m_i=m$, corresponding to the $\alpha$-particle mass.
Several definitions of the Jacobi coordinates exist, but a convenient one used throughout this work is
\begin{equation}
\boldsymbol{x}_{N-j+1}=\sqrt{\frac{2 j}{j+1}}\left(\boldsymbol{r}_{j+1}-\frac{1}{j} \sum_{i=1}^j \boldsymbol{r}_i\right) .  \label{eq:jacobicoordinate}
\end{equation}

From a given choice of the Jacobi vectors, the hyperspherical coordinates $\left(\rho, \Omega_N\right)$ can be introduced. The hyperradius $\rho$ is defined by
\begin{align}
\rho=\sqrt{\sum_{i=1}^N x_i^2}&=\sqrt{\frac{2}{A} \sum_{j>i=1}^A\left(\boldsymbol{r}_i-\boldsymbol{r}_j\right)^2}\nonumber\\&=\sqrt{2 \sum_{i=1}^A\left(\boldsymbol{r}_i-\boldsymbol{X}\right)^2},\label{eq:rho}
\end{align}
where $x_i$ is the modulus of the Jacobi vector $\boldsymbol{x}_i$.
The set $\Omega_N$ of hyperangular coordinates,
\begin{equation}
\Omega_N=\left\{\hat{\bmx}_1, \ldots, \hat{\boldsymbol{x}}_N, \varphi_2, \ldots, \varphi_N\right\},\label{eq:omega}
\end{equation}
is made of the angular parts $\hat{\boldsymbol{x}}_i=\left(\theta_i, \phi_i\right)$ of the spherical components of the Jacobi vectors $\boldsymbol{x}_i$, and of the hyperangles $\varphi_i$, defined by
\begin{equation}
\cos \varphi_i=\frac{x_i}{\sqrt{x_1^2+\ldots+x_i^2}},\label{eq:cos}
\end{equation}
where $0 \leq \varphi_i \leq \pi / 2$ and $i=2, \ldots, N$.
The advantage of using the hyperspherical coordinates can be appreciated by noting that the internal kinetic energy operator of the $A$-body system can be decomposed as
\begin{align}
T&=-\frac{\hbar^2}{m} \sum_{i=1}^N \Delta_{\bmx_i}\nonumber\\&=-\frac{\hbar^2}{m}\left(\frac{\partial^2}{\partial \rho^2}+\frac{3 N-1}{\rho} \frac{\partial}{\partial \rho}-\frac{\Lambda_N^2\left(\Omega_N\right)}{\rho^2}\right),
\end{align}
where the operator $\Lambda_N^2\left(\Omega_N\right)$ is the so-called grand-angular momentum operator. The eigenfunctions of the grand-angular momentum $\Lambda_N^2\left(\Omega_N\right)$, the so-called HH functions, can be defined as
\begin{align}
\mathcal{Y}_{[K]}^{K L M_L}\left(\Omega_N\right)&=[[\left[Y_{l_1}\left(\hat{\boldsymbol{x}}_1\right) Y_{l_2}\left(\hat{\boldsymbol{x}}_2\right)\right]_{L_2} \nonumber\\&\ldots Y_{l_{N-1}}\left(\hat{\bmx}_{N-1}\right)]_{L_{N-1}} Y_{l_N}\left(\hat{\bmx}_N\right)]_{L M_L} \nonumber\\\prod_{j=2,N}&{ }^{(j)} \mathcal{P}_{n_j}^{K_{j-1}, l_j}\left(\varphi_j\right) .\label{eq:YKLM}
\end{align}
Here $Y_{l_i}\left(\hat{\boldsymbol{x}}_i\right)$ is a spherical harmonic function for $i=1, \ldots, N$, $L$ is the total orbital angular momentum, $M_L$ its projection on the $z$ axis, and
\begin{equation}
K_j=\sum_{i=1}^j\left(l_i+2 n_i\right)\label{eq:K}
\end{equation}
with $n_1=0, j=1, \ldots, N$, and $K_N \equiv K$ being the so-called grand-angular momentum. The notation $[K]$, in Eq. (\ref{eq:YKLM}), stands for the collection of all the quantum numbers $\left[l_1, \ldots, l_N, L_2, \ldots, L_{N-1}, n_2, \ldots, n_N\right]$. The functions ${ }^{(j)} \mathcal{P}_{n_j}^{K_{j-1}, l_j}\left(\varphi_j\right)$ are defined by
\begin{align}
{ }^{(j)} \mathcal{P}_{n_j}^{K_{j-1}, l_j}\left(\varphi_j\right) =&\mathcal{N}_{n_j}^{l_j, \nu_j}\left(\cos \varphi_j\right)^{l_j}\left(\sin \varphi_j\right)^{K_{j-1}}\nonumber\\\times& P_{n_j}^{\nu_{j-1}, l_j+1 / 2}\left(\cos 2 \varphi_j\right),
\end{align}
where $P_{n_j}^{\nu_{j-1}, l_j+1 / 2}\left(\cos 2 \varphi_j\right)$ are Jacobi polynomials \cite{abramowitz+stegun}, with $\nu_j=K_j+\frac{3}{2} j-1$, and the normalization factors $\mathcal{N}_n^{l, \nu}$ are given by
\begin{equation}
\mathcal{N}_n^{l, \nu}=\sqrt{\frac{2 \nu \Gamma(\nu-n) \Gamma(n+1)}{\Gamma(\nu-n-l-1 / 2) \Gamma(n+l+3 / 2)}},
\end{equation}
$\Gamma$ indicating the standard Gamma function \cite{abramowitz+stegun}.

With the definition of Eq. (\ref{eq:YKLM}), the HH functions are eigenvectors of the grand-angular momentum operator $\Lambda_N^2\left(\Omega_N\right)$, the square of the total orbital angular momentum $L$, its $z$ component $L_z$, and the parity operator $\Pi$. 
An important property of the HH  functions is that they are orthonormal with respect to the volume element $d \Omega_N$. The number of $\mathrm{HH}$ functions increases with $K$ but is always finite. Moreover, the HH functions constitute a complete basis.

In order to evaluate matrix elements of a given many-body operator between HH functions, it is often useful to determine the effect of a particle permutation. The permuted $\mathrm{HH}$ functions $\mathcal{Y}_{[K]}^{K L M_L}\left(\Omega_N^p\right)$ can be written as linear combinations of unpermuted HH functions $\mathcal{Y}_{\left[K^{\prime}\right]}^{K L M_L}\left(\Omega_N\right)$ with the same $K, L$, and $M_L$ values \cite{Kievsky_2008}. Therefore, we can write
\begin{equation}
\mathcal{Y}_{[K]}^{K L M_L}\left(\Omega_N^p\right)=\sum_{\left[K^{\prime}\right]} a_{[K] ;\left[K^{\prime}\right]}^{K L, p} \mathcal{Y}_{\left[K^{\prime}\right]}^{K L M_L}\left(\Omega_N\right) .  \label{eq:YKL}
\end{equation}
The transformation coefficients $a_{[K]:\left[K^{\prime}\right]}^{K L, p}$ do not depend on the quantum number $M_L$. For $A=3$, they are called the Raynal-Revai coefficients \cite{RR}. Note that $\left[K^{\prime}\right] \equiv\left[l_1^{\prime}, \ldots, l_{N,}^{\prime}, L_2^{\prime}, \ldots, L_{N-1}^{\prime}, n_2^{\prime}, \ldots, n_N^{\prime}\right]$, but such that $K^{\prime}=K$. Also, $L=L$, i.e., $L$ is conserved. For $A>3$, see Ref. \cite{Kievsky_2008}.

\subsubsection{The $A=3$ and $4$ bound states}
The wave function of an $A$-boson bound state, with $A=3,4$, having total angular momentum $L, L_z$, and parity $\pi$, can be decomposed as a sum of amplitudes as
\begin{equation}
\Psi_A=\sum_{p=1}^{N_p} \psi\left(\boldsymbol{x}_1^{(p)}, \cdots, \boldsymbol{x}_N^{(p)}\right)\,.\label{eq:decpsi}
\end{equation}
Here the sum on $p$ runs up to $N_p=3$ or 12 even permutations of the $A$ clusters, for $A=3$ or $4$ respectively, and the coordinates $\boldsymbol{x}_1^{(p)}, \cdots, \boldsymbol{x}_N^{(p)}$ are the Jacobi coordinates as defined in Eq. (\ref{eq:jacobicoordinate}). It should be noted that, increasing the number of particles, different arrangements of them in sub-clusters allow for different definitions of the Jacobi coordinates. However, the HH functions defined in different sets of Jacobi coordinates result to be linearly dependent. In the following, we always refer to the set defined in Eq. (\ref{eq:jacobicoordinate}).

The coordinate-space hyperspherical coordinates are given in Eqs. (\ref{eq:rho})-(\ref{eq:cos}), and the hyperangular variables are $\varphi_2$ for $A=3$ and $\varphi_2, \varphi_3$ for $A=4$.
We rewrite here the $\mathrm{HH}$ basis of Eq. (\ref{eq:YKL}) for the $A=3$ and $4$ case, introducing the notation
\begin{align}
& {[\alpha]=\left[l_{1 \alpha}, l_{2 \alpha}, L_\alpha\right] ; \quad A=3} \nonumber\\
& {[\alpha]=\left[l_{1 \alpha}, l_{2 \alpha}, l_{3 \alpha}, L_{2 \alpha}, L_\alpha\right] ; \quad A=4}
\end{align}
so that we can write
\begin{align}
\mathcal{Y}_{[\a] n_2}^K\left(\Omega_N\right)= & {\left[Y_{l_{1 \a}}\left(\hat{\boldsymbol{x}}_1\right) Y_{l_{2 \a}}\left(\hat{\boldsymbol{x}}_2\right)\right]_{L_{\mathrm{\a}}}} \nonumber\\
&\mathcal{P}_{n_2}^{l_{1 \a}, l_{1 \a}}\left(\varphi_2\right), \label{eq:Yan2}
\end{align}
for $A=3$, and
\begin{align}
& \mathcal{Y}_{[\alpha] n_2 n_3}^K\left(\Omega_N\right)=\left[\left[Y_{l_{1 \a}}\left(\hat{\boldsymbol{x}}_1\right) Y_{l_{1 \a}}\left(\hat{\boldsymbol{x}}_2\right)\right]_{L_{2 \a}} Y_{l_{1 \a}}\left(\hat{\boldsymbol{x}}_3\right)\right]_{L_\a} \nonumber\\
& { }^{(2)} \mathcal{P}_{n_2}^{l_{1 n}, l_{1 \alpha}}\left(\varphi_2\right)^{(3)} \mathcal{P}_{n_3}^{2 n_2+l_{1 \alpha}+l_{2 \alpha}, l_{3 \alpha}}\left(\varphi_3\right) ,\label{eq:Yan3}
\end{align}
for $A=4$.
We will refer to the component with a given set $\a$ as a "channel" hereafter. 
To be noticed that, in order to ensure the symmetry of the wave function,  $l_{2 \alpha}$ for $A=3$ and $l_{3 \alpha}$ for $A=4$ must be even. Furthermore, $l_{1 \alpha}+l_{2 \alpha}$ for $A=3$ and $l_{1 \alpha}+l_{2 \alpha}+l_{3 \alpha}$ for $A=4$ must be an even or odd number corresponding to a positive or negative parity state. Even with these restrictions, there is an infinite number of channels. However, the contributions of the channels with increasing values of $l_{1 \alpha}+l_{2 \alpha}$ for $A=3$ and $l_{1 \alpha}+l_{2 \alpha}+l_{3 \alpha}$ for $A=4$ should become less important, due to the centrifugal barrier. Therefore, it is found that the number of channels with a significant contribution is relatively small. 
Using Eqs. (\ref{eq:Yan2}) and (\ref{eq:Yan3}), the $A$-cluster wave function $\Psi_A$ can be written as
\begin{equation}
\Psi_A=\sum_{\alpha, n_2} u_{\a n_2}(\rho) \sum_p \mathcal{Y}_{[\alpha] n_2}^K\left(\Omega_2^{(p)}\right), \label{eq:psia} 
\end{equation}
for $A=3$, and
\begin{equation}
\Psi_A=\sum_{\alpha, n_2, n_3} u_{\alpha n_2 n_3}(\rho) \sum_p \mathcal{Y}_{[\alpha] n_2 n_3}^K\left(\Omega_3^{(p)}\right), \label{eq:psia2}  
\end{equation}
for $A=4$. The sum over $n_2$ in Eq. (\ref{eq:psia}) and $n_2, n_3$ in Eq. (\ref{eq:psia2}) is restricted to independent states. The hyperradial functions $u_{\a n_2}(\rho)$ and $u_{\a n_2 n_3}(\rho)$ are themselves expanded in terms of Laguerre polynomials, i.e. 
\begin{equation}
u_{\alpha n_2 / \alpha n_2 n_3}(\rho)=\sum_m c_{\alpha n_2 / \alpha n_2 n_3 ; m}\, f_m(\rho),
\end{equation}
where the functions $f_m(\rho)$ are written as
\begin{equation}
f_m(\rho)=\gamma^{D / 2} \sqrt{\frac{m !}{(m+D-1) !}} L_m^{(D-1)}(\gamma \rho) \mathrm{e}^{-\gamma \rho / 2} .\label{Laguerre}
\end{equation}
Here $D \equiv 3 N-1, L_m^{(D-1)}(\gamma \rho)$ is a Laguerre polynomial \cite{abramowitz+stegun}, and $\gamma$ is a non-linear parameter, to be variationally optimized. The exponential factor $\mathrm{e}^{-\gamma \rho / 2}$ ensures that $f_m(\rho) \rightarrow 0$ for $\rho \rightarrow \infty$. The optimal value of $\gamma$ depends on the potential model, and it is  in the interval $2.5-4.5\, \mathrm{fm}^{-1}$ for the study presented here.

At the end, the $A$-cluster wave function can be cast in the form
\begin{equation}
\Psi_A=\sum_{K, m} c_{K ; m}|K, m\rangle\label{eq:psidec}
\end{equation}
where
\begin{equation}
|K, m\rangle \equiv f_m(\rho) \sum_p \mathcal{Y}_{[\a] n_2 /[\alpha] n_2 n_3}^K\left(\Omega_N^{(p)}\right).
\end{equation}

Once the symmetric bosonic  state $|K, m\rangle$ is constructed, what is left is to obtain the unknown coefficients $c_{K ; m}$ of the expansion. In order to do so, we apply the Rayleigh-Ritz variational principle, which requires  the quantity $\left\langle\Psi_A|H-E| \Psi_A\right\rangle$ to be stationary with respect to the variation of any unknown coefficient. Here $H$ is the  Hamiltonian and $E$ the energy of the state.
Differentiating respect to the coefficients $c_{K ; m}$, we obtain 
\begin{align}
\sum_{K^{\prime}, m^{\prime}}\langle K, m|H&| K^{\prime}, m^{\prime}\rangle c_{K^{\prime} ; m^{\prime}}\nonumber\\=&E \sum_{K^{\prime}, m^r}\left\langle K, m|\mathbb{1}| K^{\prime}, m^{\prime}\right\rangle c_{K^{\prime} ; m^{\prime}},  \label{eq:eigenvalue}  
\end{align}
where the matrix elements of the Hamiltonian $H$ and of the identity operator $\mathbb{1}$ can be calculated with standard numerical techniques, (see Refs.\cite{Kievsky_2008,10.3389/fphy.2020.00069} and references therein). Eq. (\ref{eq:eigenvalue}) represents a generalized eigenvalue-eigenvector problem, which can be solved with a variety of numerical algorithms.

%%%%%%%%%%
\section{Results}\label{sec:results}
\subsection{The EFT potential}
To determine the LECs of the EFT $\a$-$\a$ potential in Eq.(\ref{eq:Veff}), we performed a fit on the $S$- and $D$-wave $\a$-$\a$ scattering data of Ref. \cite{expdata} in the  low-energy regime, i.e. for $E_{CM}\leq5$ MeV.
The two-body scattering state is solved using the  Kohn variational principle \cite{KIEVSKYKOHN}.
In this case the wave function, $\Psi_{LL_z}$ is written as
\begin{equation}
\Psi_{LL_z}=\Psi_{core}+\Psi_{asym}\,,
\end{equation}
where $\Psi_{asym}$ is the  asymptotic wave function,
\begin{equation} 
\Psi_{asym}=\Omega_{L L_{z}}^{-}+
 \mathcal{S}_{L , L}(q) \Omega_{L L_z}^{+}.\label{eq:PsiA}
\end{equation}
Here $\mathcal{S}_{L , L}(q)$  are the $S$-matrix elements, $q$ the $\a-\a$ relative momentum and $\Omega_{L L_z}^{\pm}$ represent linear combinations of the regular and irregular solutions of the two-body Schr\"odinger equation, properly regularized,  with Coulomb potential. The function $\Psi_{core}$ describes the system when the   $\a$  particles  are close to each other, it  is again  decomposed similarly to Eq. (\ref{eq:psidec}). 
From the $S$-matrix it is possible to compute the phase shifts, from which the scattering observables are obtained. The $S$-matrix in Eq.~(\ref{eq:PsiA}) and  the coefficients $c_{K;m}$ in Eq. (\ref{eq:psidec}) are obtained applying the complex Kohn variational principle, i.e.requiring that the functional 
\begin{equation}
\left[\mathcal{S}_{L , L }(q)\right]=\mathcal{S}_{L , L }(q)-\frac{i}{2} \left\langle  \Psi_{L  L_z}|H-E| \Psi_{L  L_z}\right\rangle\label{eq:kohnvar}
\end{equation}
be stationary under variations of the trial parameters in $\Psi_{L  L_z}$.
This implies that the weight $\mathcal{S}_{L , L}(q)$ must solve the equation
\begin{equation}\label{eq:systm}
\mathcal{S}_{L ,L} X_{L , L}=Y_{L , L }
\end{equation}
with
\begin{eqnarray}
X_{L , L}&=&\left\langle\Omega_{L  L_z}^{+}|H-E| \Psi_{core}^{+} +\Omega_{L  L_z}^{+}\right\rangle, \label{eq:Xls}
\\ 
Y_{L , L}&=&\left\langle\Omega_{L L_z}^{-}+\Psi_{core}^{-}|E- H| \Omega_{L L_z}^{+}\right\rangle. \label{eq:Yls}
\end{eqnarray}
Here the functions $\Psi_{core}^{\pm}$ are given in Eq. (\ref{eq:psidec}) with the coefficients $c_{K;m}^{\pm}$ being the solutions of

\begin{eqnarray}
&&\sum_{K',m'}\left\langle K;m|H-E| K';m^{\prime}\right\rangle c_{K';m'}^{\pm}=\nonumber\\&&-\left\langle K;m|H-E| \Omega_{L   L_z}^{\pm}\right\rangle\,.\label{eq:cmu}
\end{eqnarray}
Substituting the calculated weights $\mathcal{S}_{L , L }$ of Eq. (\ref{eq:systm}) into Eq. ( \ref{eq:kohnvar}), it is possible to obtain a second order estimate.

In order to fit the LECs,  we used the routine of minimization  MIGRAD which belongs to the Minuit routine of the Cern library \cite{minuit}.
The fit strategy consists of two steps.  First, for a fixed cutoff of the highest order terms, one looks for the LEC of the LO by trying to reproduce the position of the resonance of the $\beo$.  Next, one tries to reproduce the $S$-wave and $D$-wave experimental phases   and the width of the resonance using the other LECs and the cutoff of the LO term.
Through the use of two different cutoffs, $\Lambda_0$ for the leading term (which encodes the resonance energy scale) and  $\Lambda$ for higher order terms, we have been  able to obtain a complete and accurate description of the experimental data.

\begin{figure}
    \centering
    \includegraphics[scale=0.35]{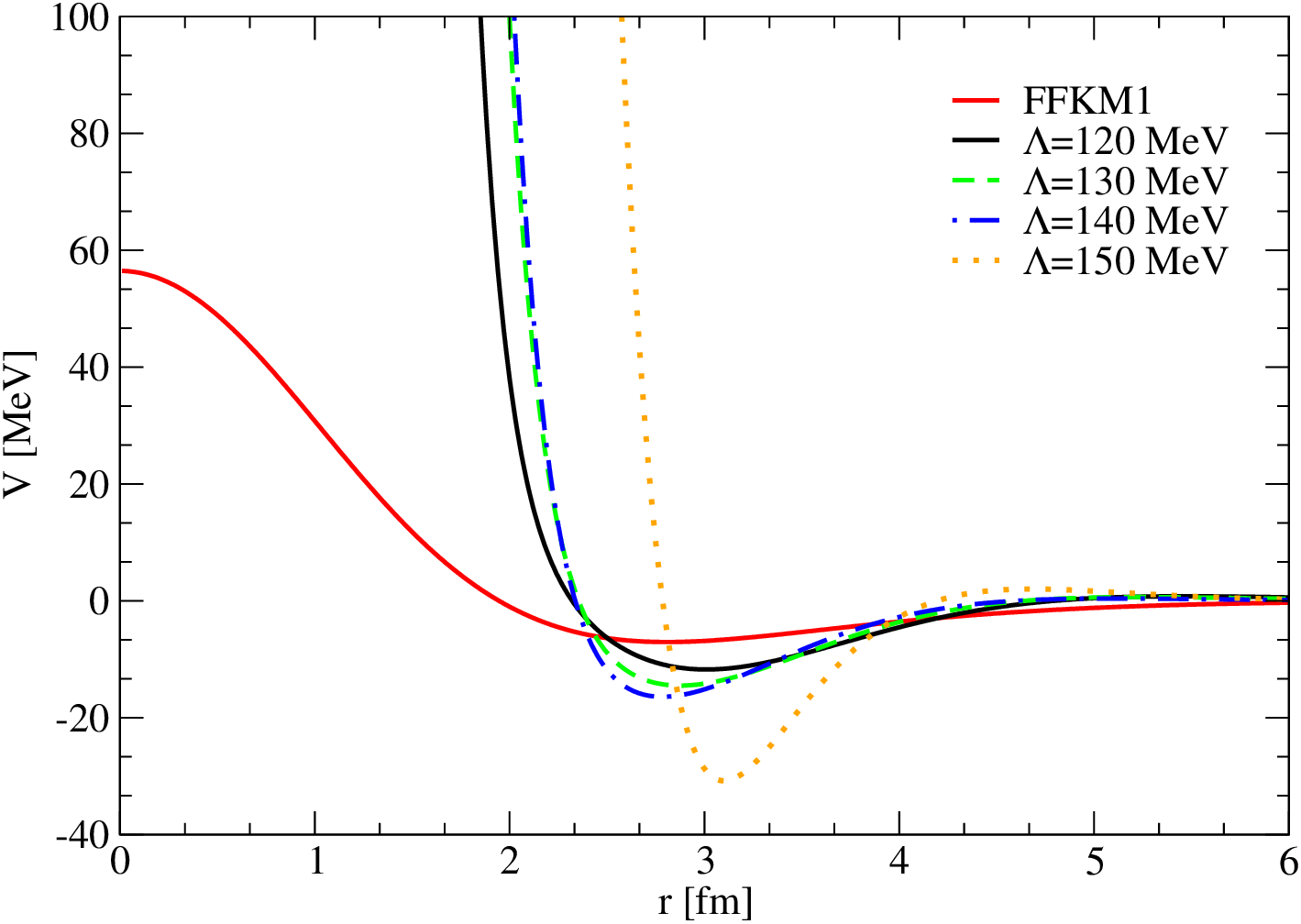}
    \caption{The LO  $\a$-$\a$ potential as a  function of the relative distance $r$, for different  values of the cutoff $\Lambda$, in comparison with the  FFKM1 potential of Ref. \cite{PhysRevC.70.014006}.}
    \label{fig:lopot}
\end{figure}

\begin{figure}
    \centering
    \includegraphics[scale=0.35]{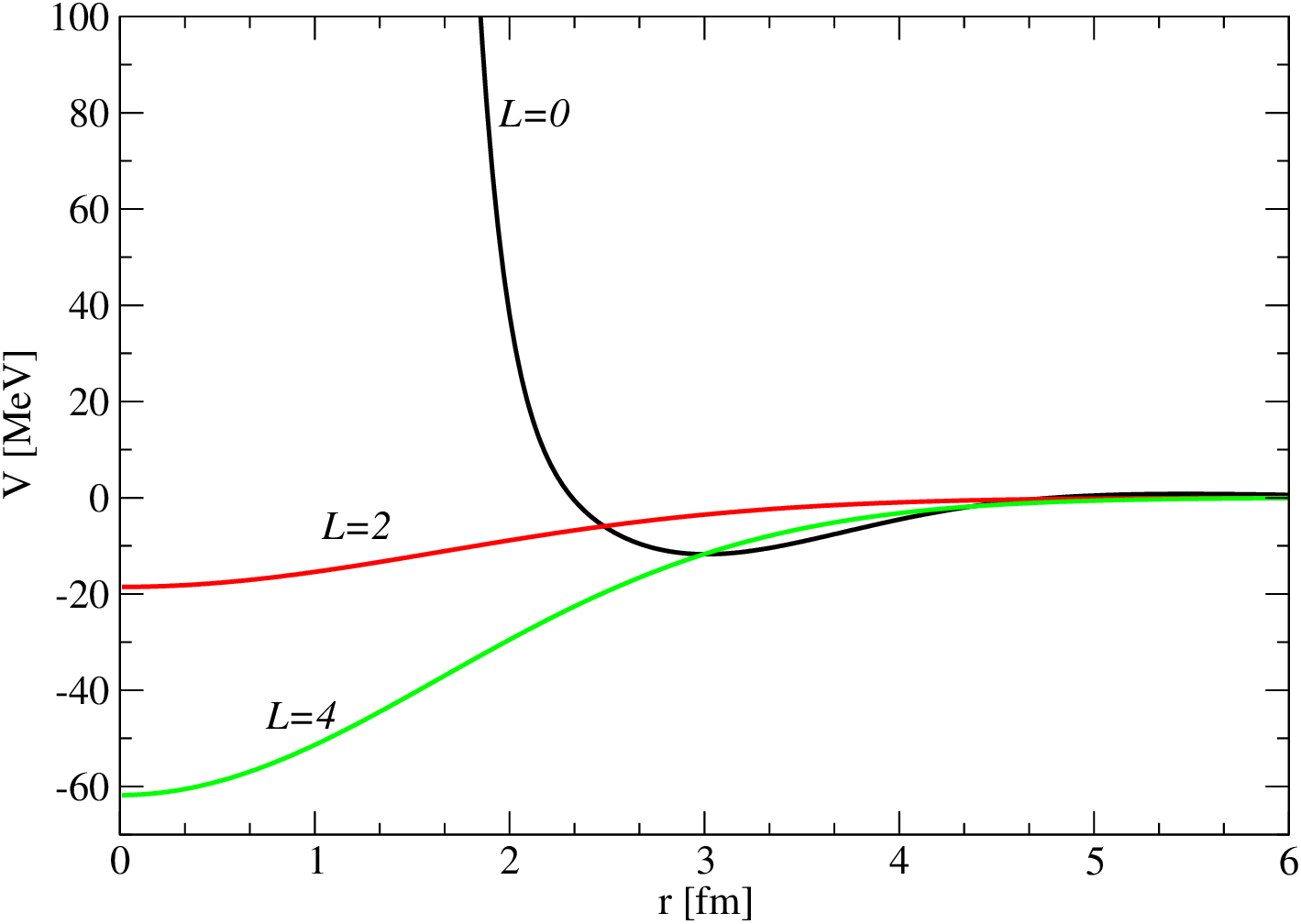}
    \caption{Partial wave decomposition of the LO $\a$-$\a$ potential for the cutoff value $\Lambda=120$ MeV, as a function of the $\a$-$\a$ relative distance $r$.}
    \label{fig:Lpot}
\end{figure}

In Table \ref{tab:Lecs} the LEC and $\Lambda_0$ values for different $\Lambda$ cutoffs as achieved from the fit are shown.
Using a total of 28 experimental data ($14$ for the $S$- and $14$ for the $D$-wave) we obtain a  $\chi^2/datum$ which is around one for each cutoff value analyzed.  For example, with $\Lambda=140$ MeV  we obtain $\chi^2=20.84$, i.e. $\chi^2/datum\sim 0.74$.
\begin{table}
    \centering
\begin{tabular}{c|c|c|c|c|c}
LECs&$\Lambda=120$&$\Lambda=130$&$\Lambda=140$&$\Lambda=150$&$\Lambda=160$\\
\hline
$C_1$&61171.417&59459.301&60631.730 &64520.692&77335.834\\
\hline
$C_2$& -3.318 & -13.884 &-29.350 &-101.4158& -614.386\\
\hline
$C_3$& -3.089&-4.218&-5.834 & -9.619&-17.317 \\
\hline
$C_4$& 8.453& 9.933& 10.361 & 43.485& 173.738\\
\hline
$\Lambda_0$&388.789& 355.706&350.332 & 264.395&228.684\\
\hline
$\chi^2$& 29.275&19.587&20.838& 22.180& 29.147
    \end{tabular}
    \caption{Values for the LECs and the cutoff $\Lambda_0$ all given in MeV, obtained from the fitting procedure described in the text. The adopted values for $\Lambda$ are $120,130,140,150$ and $160$ MeV. Also the  $\chi^2$ value is shown. }
    \label{tab:Lecs}
\end{table}
The  non-natural values for $\Lambda_0$ and for the LO  constant $C_1$ can be explained as a difficulty of the theory to describe the resonance correctly.
Increasing the values of $\Lambda$, the parameter $\Lambda_0$ has a decreasing trend starting from a value of $\sim 400$ MeV and arriving for $\Lambda=160$ MeV to about $200$ MeV.
On the other hand, the values of the LECs, as $\Lambda$ increases, show an increasing trend in modulus, going more and more toward non-natural scales.
The resulting LO term of the effective  potential, shown in  Fig. \ref{fig:lopot}, is extremely repulsive at short distances (repulsion reaches $60\,000$ MeV) and rapidly decreases to zero at a distance of $\sim 2$ fm. Its form is quite different from the FFKM1 potential of Ref. \cite{PhysRevC.70.014006}, also shown in Fig. \ref{fig:lopot} for comparison. In Fig \ref{fig:Lpot}, we show the $\a$-$\a$ potential decomposed into partial waves, for $\Lambda=120$ MeV, as a  function of the relative distance $r$. As it can be seen, the strong repulsion of the potential is due to the component with relative angular momentum $L=0$, which creates a kind of short-range barrier for the two $\a$ particles.

\begin{figure*}[h]
\begin{center}
\includegraphics[scale=0.55]{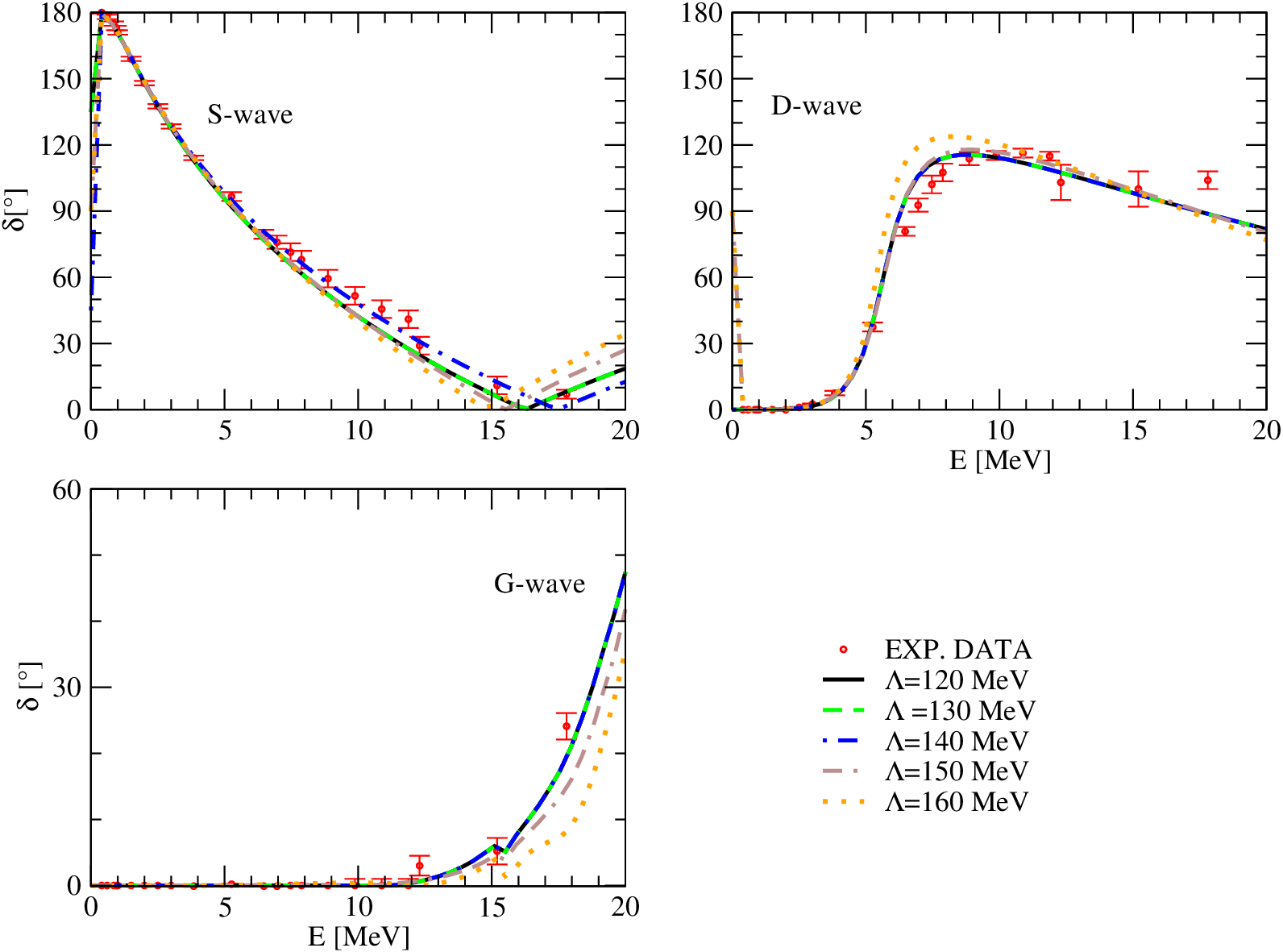}
\caption{Phase-shifts of the $\a$-$\a$ scattering process for $L=0,2,4$ for different values of the cutoff $\Lambda$. The black line represents the EFT two-body potential with $\Lambda=120$ MeV, the  dashed green one with $\Lambda=130$, the blue dotted-and-dashed line with $\Lambda=140$ MeV, the brown dotted-and-dashed line with $\Lambda=140$ MeV and the orange dotted one with $\Lambda=160$ MeV. The energy is given in the laboratory reference frame. The experimental data are taken from Ref. \cite{expdata}}\label{fig:phases}
\end{center}
\end{figure*}

The results for the  phase shifts  compared with experimental data are shown in  Fig. \ref{fig:phases}. The energy range of interest for our model in the laboratory reference frame is below $5$ MeV, and for such energies we reach a very good agreement with the data.
Note that, although only the experimental data of $S$- and $D$-waves at low energies have been fit, we manage to reproduce even higher energies up to $\sim 15$ MeV and also to predict the trend of the $G$-wave ($L=4$) without including  the relevant experimental data in the fit.
The $L(L+1)$ term in the potential is reduced to act only up to the two-body orbital angular momentum $L=4$. In fact, for larger values of $L$ it would give unnatural attractive potentials. 

\subsection{The $A=3$ system: the $\C$ nucleus}
The first nucleus we analyze is $\C$ described as  $3$-$\a$  system.
The expected ground state energy in the three-body model is $-7.275$ MeV. The ground-state energy of the three $\a$ particles, of $-28.29$ MeV each, must be added in order to reach the experimental energy of $-92.16$ MeV.
The full  interaction used in the calculation, the two-body strong  EFT potential plus the screened Coulomb repulsive interaction, is given in Eq. (\ref{eq:Vtot}) and the LECs for different $\Lambda$ values are listed in Table \ref{tab:Lecs}.
The first step to perform the calculation is to choose a HH basis large enough to achieve convergence. Due to the strong repulsion present in the LO term of the two-body potential (see Fig.\ref{fig:lopot}), the convergence requires to include HH functions with large values of grand-angular momentum  K (up to 50), and a large number of Laguerre polynomials in the expansion of the hyperradial functions.

In the following, we introduce the quantity $\mathcal{L}=\sum_{i=1,N} l_i$. Usually, we include in the expansion basis all HH functions with a given value of $\mathcal{L}$ up to a grand angular quantum number $K(\mathcal{L})$, and study the convergence by increasing such a quantity. We start from the lowest value of $\mathcal{L}$ as possible, usually $\mathcal{L}=0$ or $2$, and then include states with larger and larger values of $\mathcal{L}$. The idea is that for increasing values of $\mathcal{L}$, the corresponding HH should be contributing less and less, due to the increase of the centrifugal barrier. Note that, due to the parity constraint, in this work we need to consider only even values of $\mathcal{L}$. For example, for $A=3$, $\mathcal{L}=l_1+l_2$, and we have included states up to $\mathcal{L}=8$.

Using a basis set of  $M=44$ Laguerre polynomials,  grand-angular momentum $K(\mathcal{L})=48$ for each considered $\mathcal{L}=0,\ldots,8$,  and the variational parameter $\gamma=3$ fm$^{-1}$ (see Eq.(\ref{Laguerre})), we obtain stability in the third  decimal digit.
The results obtained using only two-body interactions are shown in Table \ref{tab:2B-12C} and compared with experimental data.
\begin{table}[hbt]
    \centering
    \begin{tabular}{c|c|c|c|c|c|c}
 $J^{\pi}$&$\Lambda=120$&$\Lambda=130$&$\Lambda=140$&$\Lambda=150$&$\Lambda=160$&EXP.\\
            \hline   
        $0^+$ &-1.803&-2.060&-2.223&-3.191 &-5.605 &-7.275 \\        
        $2^+$ &0.612&0.360& 0.143&-0.556&-3.136 &-2.875 \\
         $0^+_2$ &0.787&0.788& 0.788 &0.789&-0.108 &0.380 \\
    \end{tabular}
    \caption{Energy of $\C$ states (in MeV), using  the two-body potentials with different $\Lambda$ values, in MeV.}
    \label{tab:2B-12C}
\end{table}
As it can be seen, for each value of $\Lambda$, the binding energy of the ground state turns out to be less than the expected $7.275$ MeV.
This result is  not dissimilar to other calculations that attempt to describe $\C$ using micro-clustering models, see for example Ref.~\cite{Ronen:2005xs}.
The results range from a minimum of $-1.8$ MeV for $\Lambda=120$ MeV to a maximum of $-5.60$ MeV obtained with $\Lambda=160$ MeV.
Regarding  the $2^+$ and $0^+_2$ (Hoyle state) excited states  the obtained  energies are lower for all cutoffs except $\Lambda=160$ MeV.
To bridge  the gap in energies between theoretical predictions and experimental data, as required by EFT power counting, we include in our LO model the three-body force of  Eq. (\ref{3bforce}), tuning the $V_{03},V_{23}$ and $a_3$  parameters on the  energies of the ground state, the Hoyle state and the $2^+$ excited state, respectively.
With this procedure, we obtain the results reported in Table \ref{tab:3bf}.
\begin{table}[hbt]
    \centering
    \begin{tabular}{c|c|c|c|c}
     &$\Lambda=120$ &$\Lambda=130$ &$\Lambda=140$ &$\Lambda=150$ \\
     \hline
    $V_{03}$ [MeV]&$-19.44$&$-18.75$&$-18.58$&$-18.74$\\
    \hline
    $V_{23}$ [MeV]&$-12.50$&$-11.68$&$-11.04$&$-10.69$ \\
        \hline
    $a_3$ [fm]&$3.283088$&$3.222431$& $3.058219$ &$3.135534$
    \end{tabular}
    \caption{Three-$\a$ potential parameters $V_{03}$, $V_{23}$ and  $a_3$ as defined in Eq.(\ref{3bforce}), for  different  $\Lambda$ values, in MeV.}
    \label{tab:3bf}
\end{table}
For the various $\Lambda$ values  analyzed, the variations in the strength and cutoff of the three-body force turn out to be rather small, with $V_{i3},\,i=0,2$ of the order of $\sim$ 10 MeV   and $a_3$ of the order of $\sim 1$ fm. 
Regarding $\Lambda=160$ MeV, it was not possible to find a three-body force such that the ground state and the Hoyle state could be reproduced simultaneously. In fact, as it can be seen from Table \ref{tab:3bf}, while for the ground state an attractive force is required, for the Hoyle state  a repulsive interaction is needed so that the state would return above  threshold.

\begin{figure*}[h]
\begin{center}
\includegraphics[scale=0.4]{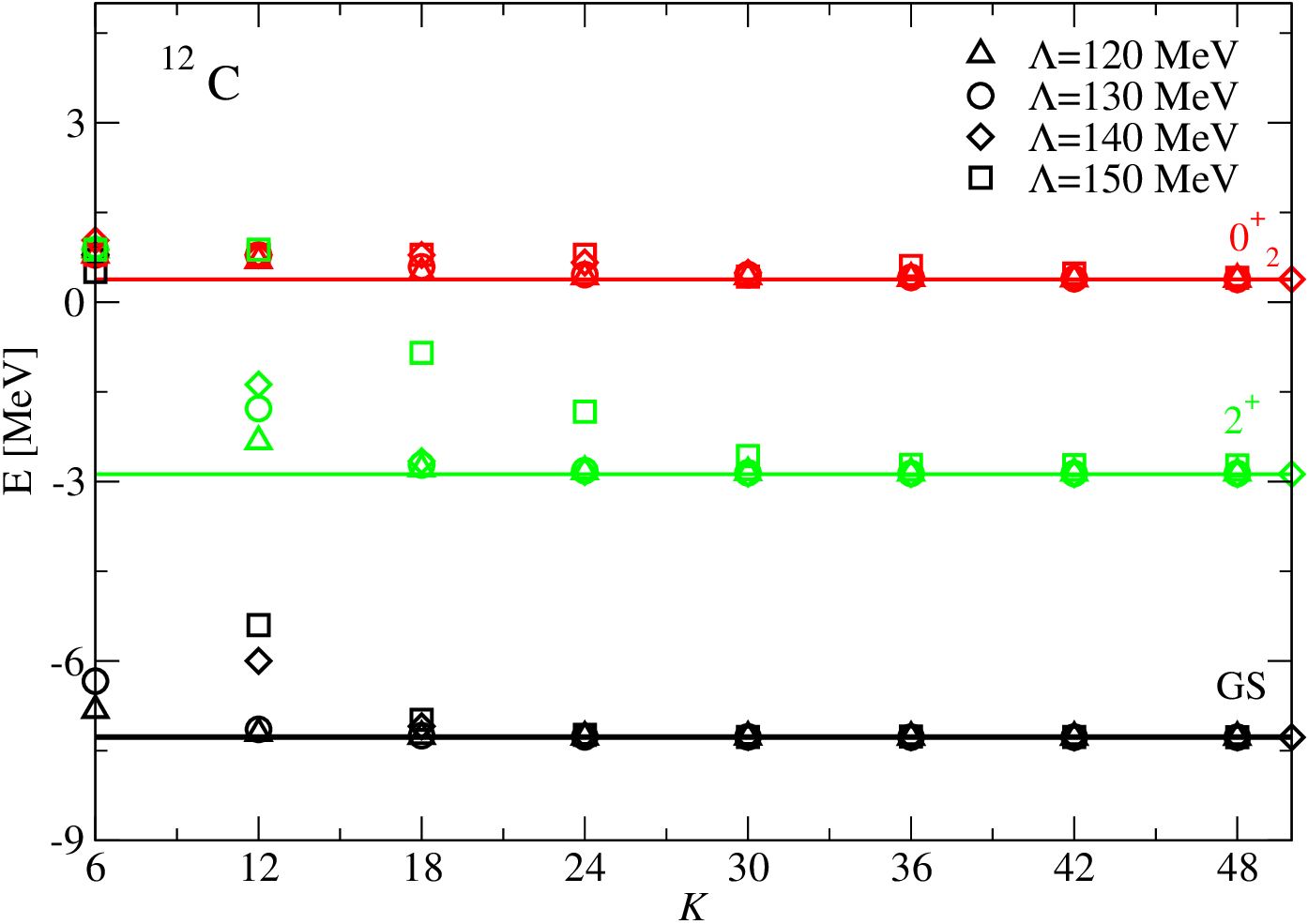}
\caption{The $\C$ ground state (black line), Hoyle state (red line), $2^+$  state (green line) energies as a function of the grand-angular momentum $K$, for different values of the cutoff $\Lambda$. The energy for a given value of $K$ has been obtained including all HH states with maximum grand angular quantum number $K(\mathcal{L})=K$, $\mathcal{L}=0,\ldots,8$.}\label{fig:CKconv}
\end{center}
\end{figure*}

In Fig. \ref{fig:CKconv} we show the convergence of the $\C$ ground and excited  state energies, as a function of the maximum grand-angular momentum values $K(\mathcal{L})$, considered the same for each value $\mathcal{L}=0,2,4,6,8$. In Table \ref{tab:12CN} we show the ground state, $2^+$ and Hoyle state energies at increasing values of the maximum number of Laguerre polynomials $M$ adopted. As it can been seen from the table, the results are very stable.
\begin{table}[hbt]
    \centering
    \begin{tabular}{c|c|c|c}
 $M$&$E_{G.S.}$ [MeV]&$E_{2^+}$ [MeV]&$E_{Hoyle}$ [MeV]\\
            \hline   
       20 &-7.1724&-2.7764&0.4068\\
        30& -7.2664&-2.8664& 0.3861\\
     40 &-7.2755&-2.8750&0.3843 \\
        50 &-7.2764&-2.8762&0.3841 \\
    \end{tabular}
    \caption{The  $\C$ ground, $2^+$ and Hoyle state energies (in MeV) as a function of the maximum number of Laguerre polynomials $M$.}
    \label{tab:12CN}
\end{table}

\subsection{The $A=4$ system: the $\Ox$ nucleus}\label{sec:Ox}
The $\Ox$ nucleus described as a $4$-$\a$  system has an experimental binding energy of $-14.437$ MeV.
As we already mentioned, the EFT with $\a$ particles as degrees of freedom is based on the assumption that the  energy of the system is much larger than the separation energy of the degree of freedom, thus $\sim 20$ MeV.
Although for $\Ox$ and the $\a$-particle the energies are comparable, the description of the $\Ox$ fundamental state within the theory is still possible, since the energy of each single $\a$ particle within the nucleus results  $-3.61$ MeV. Therefore, comparing the latter value with the excitation energy of the $\a$ particle, one can see that the assumptions for an effective theory approach could be considered still valid.

\begin{figure}[h]
\begin{center}
\includegraphics[scale=0.35]{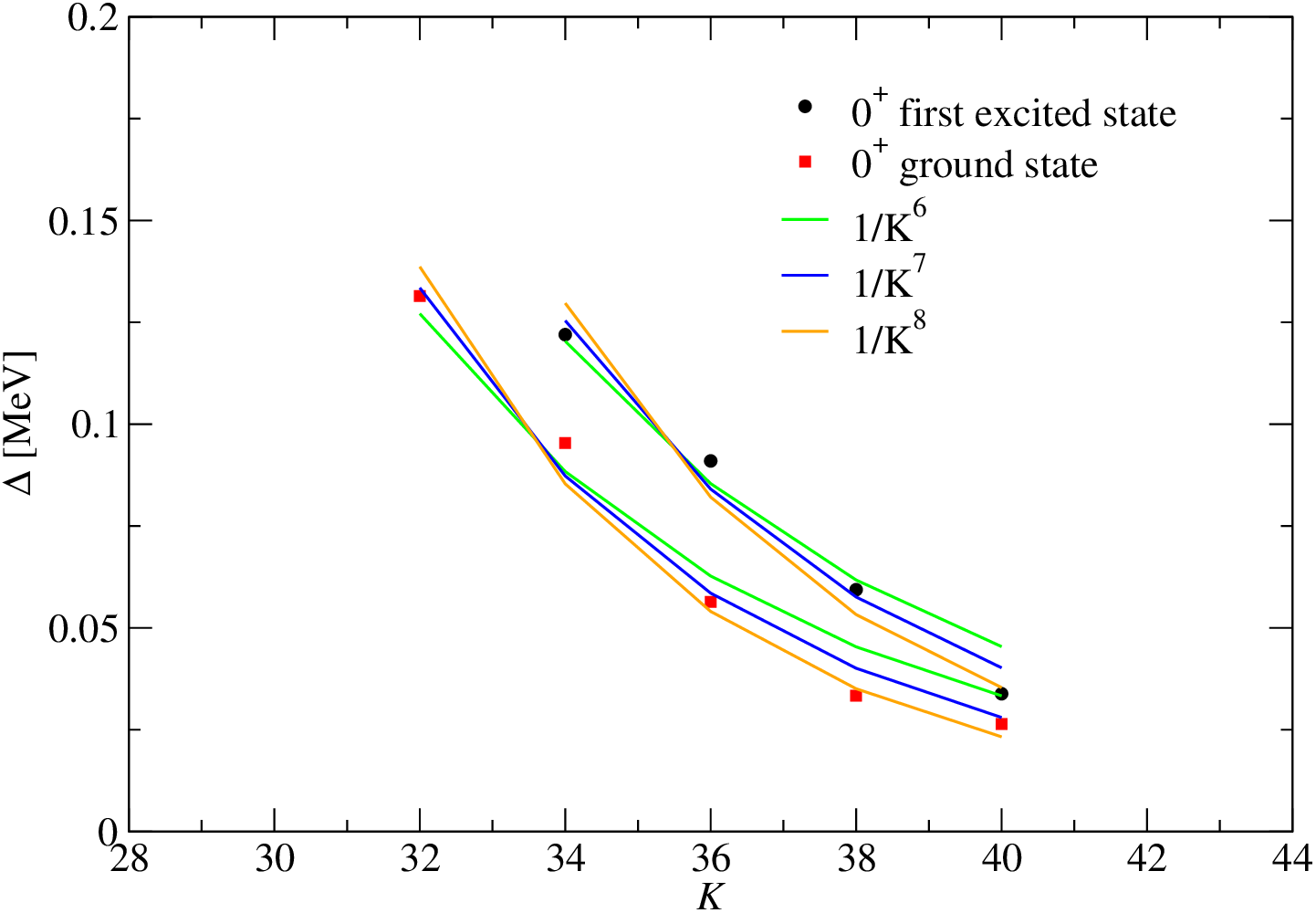}
\caption{$\Delta(K)$ for the  $\Ox$ ground state and first excited state (symbols), see text for more details.
The fit to the calculated values using different $1/K^p$ functions is also shown.}\label{fig:deltak}
\end{center}
\end{figure}

The  interaction used in the calculation  is the $\a$-$\a$ strong  EFT potential plus the screened Coulomb repulsive interaction, shown in Eq. (\ref{eq:Vtot}) with  LECs  given in Table \ref{tab:Lecs}, and the 3$\a$ potential of Eq.(\ref{3bforce}) with the parameters reported in Table \ref{tab:3bf}.
As in the case of $\C$, also for $\Ox$ the HH basis required to reach convergence is quite large, due to the fact that the two-body potential at  LO term is extremely repulsive at short distances, as shown in Fig. \ref{fig:lopot}.
From this study, we excluded the case with $\Lambda= 150,\, 160$ MeV being extremely sharp and making convergence very slow in the $4$-$\a$ case. Moreover, the non-natural scales of such cases make us lean toward potentials with smoother cutoffs.
For the $4$-$\a$ problem, being the Hamiltonian matrix of higher dimensions, it is necessary to resort to fast diagonalization procedures to solve the eigenvalue  of Eq.(\ref{eq:eigenvalue}).  We tested the convergence of the $4$-$\a$ system with different numerical procedures: the Lanczos algorithm \cite{doi:10.1137/1.9780898719192.ch2}, the Andreozzi-Porrino-Lo Iudice method \cite{Andreozzi_2003}, that solves only for a finite number of negative eigenvalues, and  the Lapack \cite{lapack99} standard subroutine dspev. The latter algorithm allows for the storage of the matrix elements of the Hamiltonian $H_{i,j=1,\ldots,N_{dim}}$,  $N_{dim}$ being the dimension determined by the expansion on the HH basis, in a vector containing  only the upper triangular part of $H_{ij}$. Therefore  we can save  half memory space, so the size of the matrix can be increased by $\sqrt{2}$. 
In addition, to check the stability of each result obtained, we used the algorithm presented below to extrapolate the result at convergence with increasing values of the grand-angular momentum $K$. 
This algorithm is based on the  estimate of the  missing  energy due to the truncation of the expansion to finite values of $K=\bar{K}$. Let us suppose that the states  up to $K=\bar{K}$ have been included and to have computed  $\Delta(K)=B(K)-B(K-2)$, for $K\to \bar{K}$, being $B(K)$ the binding energy calculated including states up to $K$. If the asymptotic behavior of $\Delta(K)$ is proportional to an inverse power $p$ of $K$,  then the  missing binding energy $\Delta B$ due to the states with $K=\bar{K}+2,\bar{K}+4,\ldots$ reads
\begin{equation}
(\Delta B)=c(\bar{K}, p) \Delta(\bar{K}), \quad c(\bar{K}, p)=\!\!\sum_{K=\bar{K}+2, \bar{K}+4, \ldots}^{\infty}(\bar{K}/K)^p,\label{eq:deltaB}
\end{equation}
where $c(\bar{K}, p)$ is a numerical coefficient. In our case,  $\Delta(K)\propto 1/K^7$ for $K\to\infty$, as it can be seen in Fig \ref{fig:deltak}, therefore we set $p=7$.

With a basis set of  $M\ge28$ Laguerre polynomials,  $\mathcal{L}=0,\ldots,6$,  maximum grand-angular momentum values $K(0)=40,\,K(2)=34,\,K(4)= 34,\,K(6)=26$, and $\gamma=3$ fm$^{-1}$ (see Eq.( \ref{Laguerre})), we obtain a quite stable result  for the ground-state  energy with an error at the percent level. We show in Table \ref{tab:0K} the convergence of the ground-state energy for $\Lambda=140$ MeV as a function of angular momentum $K(\mathcal{L})$ for each considered $\mathcal{L}$, while in Table \ref{tab:Om} is shown the convergence in the number of Laguerre polynomials $M$.
The results obtained using  two-body  and three-body interactions are shown in Table \ref{tab:16O}, in comparison with experimental data.

\begin{table}[hbt]
    \centering
    \begin{tabular}{c|c|c|c|c}
 $K(0)$&$K(2)$&$K(4)$&$K(6)$&$E_{GS}$\\
            \hline   
   30& 24& 24& 16 &    -24.98\\
     32& 26& 26& 18  &-25.39\\
    34 &28& 28& 20&    -25.64\\
   36 &30& 30 &22 &    -25.84\\
   38& 32& 32 &24 &    -26.00\\
   40 &34& 34& 26  &   -26.06
    \end{tabular}
    \caption{The $\Ox$ ground-state  energy (in MeV) as a function of the maximum grand-angular momentum values $K(\mathcal{L})$, calculated with  $M=28$ Laguerre polynomials, for the $\a$-$\a$ potential and $3\a$ potentials with $\Lambda=140$ MeV.}
    \label{tab:0K}
\end{table}

\begin{table}[hbt]
    \centering
    \begin{tabular}{c|c|c|c|c}
 $M$&$\Lambda=120$&$\Lambda=130$&$\Lambda=140$\\
            \hline   
       28 &-26.621&-26.042&-26.058\\
        30& -26.627&-26.060&-26.097\\
        32& -26.628&-26.069&-26.112\\
        34& -26.628&-26.078&-26.131\\
        36& -26.628&-26.081&-26.139
    \end{tabular}
    \caption{The $\Ox$ ground state  energy (in MeV) as a function of the maximum number of Laguerre polynomials $M$, using the $\a$-$\a$ and $3\a$ potentials with different $\Lambda$ values in MeV.}
    \label{tab:Om}
\end{table}

\begin{table}[hbt]
    \centering
    \begin{tabular}{c|c|c|c|c|c}
 $J^{\pi}$&$\Lambda=120$&$\Lambda=130$&$\Lambda=140$&EXP.\\
            \hline   
        $0^+$ &-26.675&-26.224&-26.371&-14.437 \\
        $0^+_2$ & -14.484 &-14.095&-14.203&-8.388\\
        $2^+$ &-11.968&-9.962&-11.967&-7.520 \\
    \end{tabular}
    \caption{Energies of $\Ox$ states in MeV  with the $\a$-$\a$ and $3\a$ potentials, using different $\Lambda$ values in MeV. Here $M=34$. }
    \label{tab:16O}
\end{table}

As it can be seen in Tables \ref{tab:Om} and \ref{tab:16O}, the three-body force fitted to reproduce the $\C$ states leads to over binding for the  $\Ox$  ground-state, with an energy $E_{GS}\sim - 26$ MeV against an expected experimental energy of $-14.437$ MeV.

As already  mentioned, in order to reproduce the $\Ox$  levels, we need to add the four-body force of Eq. (\ref{4bforce}) to the potential, with $V_{04},V_{24}$ and $a_4$  tuned  on the  energy of the ground state, the first excited state and the  $2^+$ excited state. With this procedure, we obtain the values reported in Table \ref{tab:4bf}.
For the fit of the 4-$\a$ potential, we implemented a code that searches the potentials strength $V_{04}$ and range $a_4$ such that both the experimental ground state and the $0^+_2$ first excited state energies  are fitted. For the $2^+$ state, we just retain the same $a_4$, and vary $V_{24}$ until
the energy of the lowest $2^+$ level is reproduced.
With this particular choice of potential parameters the second excited state $2^+$ of energy $-4.592$ MeV is automatically reproduced by the model.
As it can be seen in Table \ref{tab:4bf}, the strength $V_{04}$ decreases  by about $40$ MeV as $\Lambda$ increases, while $V_{24}$ undergoes an almost twofold variation of about $80$ MeV. This is probably due to the procedure of fitting the cutoff of the $4$-$\a$ force on the $0^+_2$ excited state, which forces the strength $V_{24}$  to variate significantly in order to reproduce the energy of the $2^+$ state.

\begin{figure*}[hbt]
\begin{center}
\includegraphics[scale=0.4]{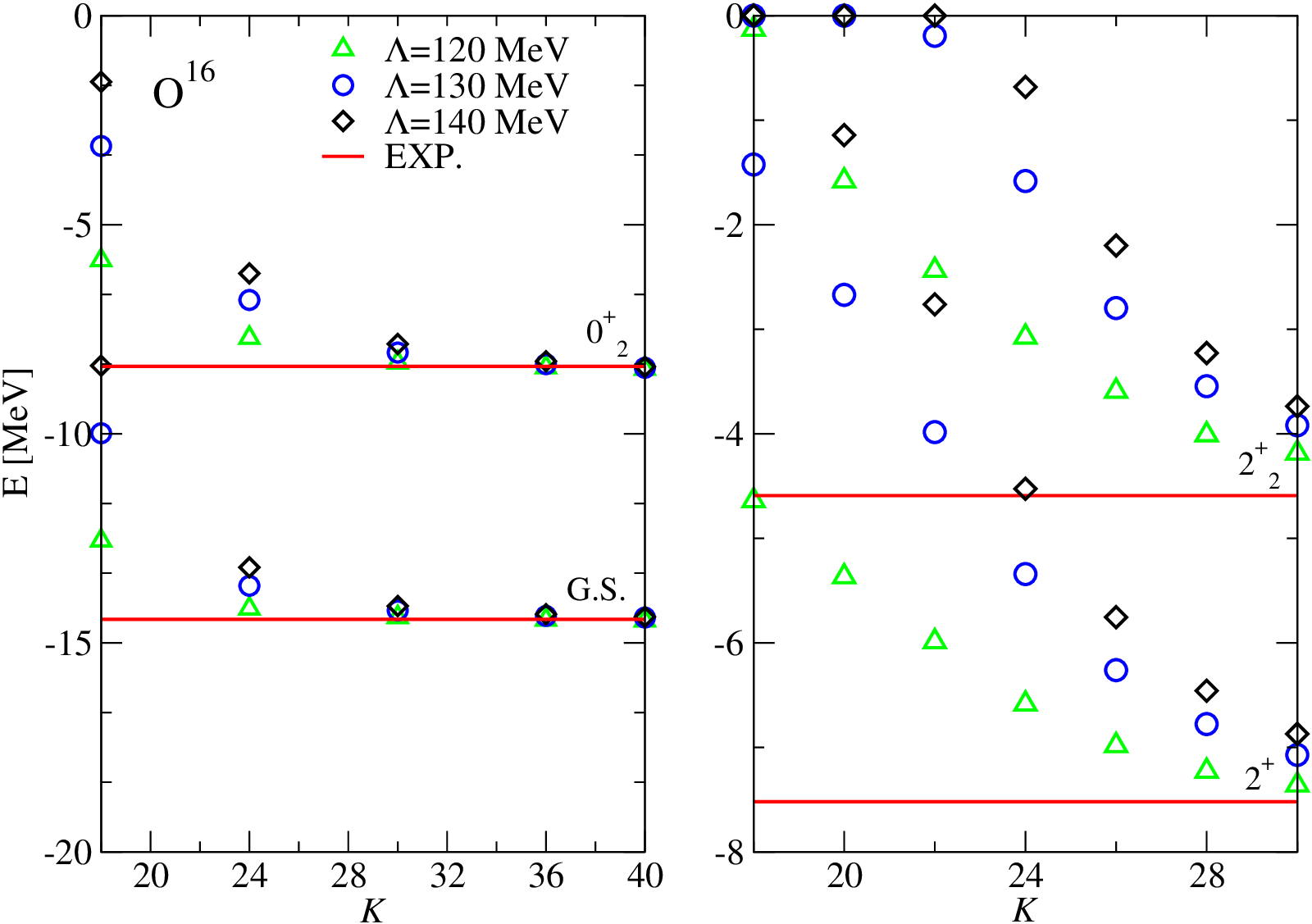}
\caption{On the left panel, the $\Ox$ ground state and  $0^+_2$  state  energies as a function of grand-angular momentum $K$. On the right panel, the $\Ox$ $2^+$  state  energies as a function of  $K$. The potential includes the $\a$-$\a$, $3\a$ and $4\a$ contributions. Various values of the cutoff $\Lambda$ are used. See text for more details.}\label{fig:OKconv}
\end{center}
\end{figure*}

\begin{table}[hbt]
    \centering
    \begin{tabular}{c|c|c|c}
    & $\Lambda=120$ &$\Lambda=130$ &$\Lambda=140$ \\
    \hline
    $V_{04}$ [MeV]&$234.41$&$207.85$&$185.10$\\
    $V_{24}$ [MeV]&$236.80$&$193.42$&$160.00$\\
    $a_4$ [fm]&$2.542874$& $2.542874$ &$2.542874$
    \end{tabular}
    \caption{Four-$\a$ potential parameters $V_{04},V_{24}$ and $a_4 $ as defined in Eq.(\ref{4bforce}) for  different  $\Lambda$ values, in MeV.}
    \label{tab:4bf}
\end{table}
We report in Fig. \ref{fig:OKconv} on the left panel the $\Ox$ ground and $0^+_2$ excited  state energies as a function of maximum grand-angular momentum values $K(\mathcal{L})$, with the inclusion of  the 4-$\a$ interaction.  Here,  the number of Laguerre polynomials used is $M=28$. The first point corresponds to $K(0)=18,\, K(2)=K(4)=16,\, K(6)=12$, the second $K(0)=24,\,K(2)=K(4)=22,\, K(6)=16$, the third $K(0)=30,\, K(2)=K(4)=28,\, K(6)=22$, the forth $K(0)=36,\, K(2)=K(4)=32 ,\, K(6)=24$ and the last $K(0)=40,\, K(2)=K(4)=34 ,\, K(6)=26$.  As it is expected, convergence is achieved at the experimental value. We also get stability of the results at the percent level as a function of the number of Lagrange polynomials $M$ as shown in Table \ref{tab:Om140}. 
In Fig.\ref{fig:OKconv} on the right panel the energies for the  $\Ox$ $2^+$ excited states are shown. Here the points correspond to: $K(2)=18,\, K(4)=14,\, K(6)=12$; $K(2)=20,\,K(4)=16,\, K(6)=14$; $K(2)=22,\, K(4)=18,\, K(6)=16$; $K(2)=24,\, K(4)=20,\, K(6)=18$; $K(2)=26,\, K(4)=22,\, K(6)=20$; $K(2)=28,\, K(4)=24,\, K(6)=22$; $K(2)=30,\, K(4)=26,\, K(6)=24$.
Due to computational power limitations we have not been able to  insert $\mathcal{L}=8$ waves and to go beyond the point $K(2)=30,\, K(4)=26,\, K(6)=24$, thus not achieving full convergence. For the first excited state $2^+$ we obtain a difference from the experimental value of $0.3$ MeV, while for the second one we obtain $0.6$ MeV.
Here, in order to obtain the values of the energies at full convergence we used the extrapolation algorithm presented in Sec. \ref{sec:Ox} (see Eq. (\ref{eq:deltaB})). Since we have observed that the convergence of HH functions with the minimum value of $\mathcal{L}\equiv\mathcal{L}_{min}$ is the most critical, in Eq. (\ref{eq:deltaB}) we have considered $K=K(\mathcal{L}_{min})$. We report the energy values obtained by the extrapolation in Table \ref{tab:Ex2+}. Note that in this case the calculation results are also stable by varying the maximum number of Laguerre polynomials $M$, as shown in Table \ref{tab:Om1202} for $\Lambda=120$ MeV.

\begin{table}[hbt]
    \centering
    \begin{tabular}{c|c|c}
 $M$&$GS$&$0^+_2$\\
            \hline   
       26 &-14.298&-8.253\\
       28& -14.315&-8.266\\
        30& -14.322&-8.272\\
    \end{tabular}
    \caption{The $\Ox$ ground state and $0^+_2$ excited state energies, in MeV, as a function of the maximum number of Laguerre polynomials $M$, using a potential model with $\a$-$\a$, $3\a$ and $4\a$ contributions, with $\Lambda=140$ MeV. Here $K(0)=36,\,K(2)=32,\,K(4)=32,\,K(6)=24$.}
    \label{tab:Om140}
\end{table}

\begin{table}[hbt]
    \centering
    \begin{tabular}{c|c|c}
 $M$&$2^+$&$2^+_2$\\
            \hline   
       26 & -7.227 &-4.003\\
       28& -7.227&-4.003\\
       30& -7.228   & -4.009\\
    \end{tabular}
    \caption{The $\Ox$ $2^+$ excited state energies, in MeV, as a function of the maximum number of Laguerre polynomials $M$, using a potential model with $\a$-$\a$, $3\a$ and $4\a$ contributions, with $\Lambda=120$ MeV. Here $K(0)=28,\,K(2)=24,\,K(4)=22$.}
    \label{tab:Om1202}
\end{table}

\begin{table}[hbt]
    \centering
    \begin{tabular}{c|c|c}
    &$2^+$ &$2^+_2$\\
    \hline
$\Lambda=120$ & $-7.557$&$-4.447$\\
$\Lambda=130$ &$-7.515$&$-4.479$\\
$\Lambda=140$ &$-7.491$&$-4.500$\\
    \end{tabular}
    \caption{Extrapolated $\Ox$ $2^+$ excited state energies in MeV for $\bar{K}=\infty$ and  for  different $\Lambda$ values in MeV. Here we use the algorithm presented in the text setting $p=7$, $\Delta(K)=0.6$, see the text for more details. }
    \label{tab:Ex2+}
\end{table}
\section{Conclusions}\label{sec:conclusion}
In this paper we report an analysis of the bound and excited states of $3\a$ and $4\a$ systems in the short-range EFT.
The theory is constructed by analyzing the independent building block allowed by the low-energy spatial symmetries of the underlying fundamental theory. Developing this procedure, we constructed a contact potential up to  N2LO, neglecting only the non-local terms. To this strong interaction is added a smeared Coulomb repulsion and a three-body contact interaction.
The LECs of the $\a$-$\a$ potential are fitted on the $\a$-$\a$ scattering  phase shifts data and resonance position and width, while the cutoff and  LECs related to the 3$\a$ force are tuned to reproduce the binding energy of the fundamental and first excited states of $\C$ . 
Applying this model to the study of the 4-$\a$ system, we obtain a fundamental state $\sim -26$ MeV for each considered cutoff values. For these reasons, we  add  to the theory a 4$\a$ force such that the energies of the ground and  first  excited states are reproduced.
This work sets the starting block in order to apply the near zero-energy EFT approach to study the $\a+\a$, $\a+\beo$, $\a+\C$ radiative captures, of paramount importance in stellar evolution. Work along this line is currently ongoing.

\section*{Acknowledgement}
E.F. and L.E.M. acknowledge the financial support of the University of Pisa, via the program "Progetto di ricerca di Ateneo PRA 2022", with the title "From stars and galaxies to nuclear fusion energy production on Earth". Furthermore, the authors acknowledge the CINECA award under the ISCRA initiative for the availability of high-performance computing resources and related support. The computational resources of the INFN-Pisa branch  are also gratefully acknowledged. 
\bibliography{bib}

\providecommand{\noopsort}[1]{}\providecommand{\singleletter}[1]{#1}
\begin{thebibliography}{23}%
\makeatletter
\providecommand \@ifxundefined [1]{%
 \@ifx{#1\undefined}
}%
\providecommand \@ifnum [1]{%
 \ifnum #1\expandafter \@firstoftwo
 \else \expandafter \@secondoftwo
 \fi
}%
\providecommand \@ifx [1]{%
 \ifx #1\expandafter \@firstoftwo
 \else \expandafter \@secondoftwo
 \fi
}%
\providecommand \natexlab [1]{#1}%
\providecommand \enquote  [1]{``#1''}%
\providecommand \bibnamefont  [1]{#1}%
\providecommand \bibfnamefont [1]{#1}%
\providecommand \citenamefont [1]{#1}%
\providecommand \href@noop [0]{\@secondoftwo}%
\providecommand \href [0]{\begingroup \@sanitize@url \@href}%
\providecommand \@href[1]{\@@startlink{#1}\@@href}%
\providecommand \@@href[1]{\endgroup#1\@@endlink}%
\providecommand \@sanitize@url [0]{\catcode `\\12\catcode `\$12\catcode `\&12\catcode `\#12\catcode `\^12\catcode `\_12\catcode `\%12\relax}%
\providecommand \@@startlink[1]{}%
\providecommand \@@endlink[0]{}%
\providecommand \url  [0]{\begingroup\@sanitize@url \@url }%
\providecommand \@url [1]{\endgroup\@href {#1}{\urlprefix }}%
\providecommand \urlprefix  [0]{URL }%
\providecommand \Eprint [0]{\href }%
\providecommand \doibase [0]{http://dx.doi.org/}%
\providecommand \selectlanguage [0]{\@gobble}%
\providecommand \bibinfo  [0]{\@secondoftwo}%
\providecommand \bibfield  [0]{\@secondoftwo}%
\providecommand \translation [1]{[#1]}%
\providecommand \BibitemOpen [0]{}%
\providecommand \bibitemStop [0]{}%
\providecommand \bibitemNoStop [0]{.\EOS\space}%
\providecommand \EOS [0]{\spacefactor3000\relax}%
\providecommand \BibitemShut  [1]{\csname bibitem#1\endcsname}%
\let\auto@bib@innerbib\@empty
%</preamble>
\bibitem [{\citenamefont {Freer}\ \emph {et~al.}(2018)\citenamefont {Freer}, \citenamefont {Horiuchi}, \citenamefont {Kanada-En’yo}, \citenamefont {Lee},\ and\ \citenamefont {Meißner}}]{Freer}%
  \BibitemOpen
  \bibfield  {author} {\bibinfo {author} {\bibfnamefont {M.}~\bibnamefont {Freer}}, \bibinfo {author} {\bibfnamefont {H.}~\bibnamefont {Horiuchi}}, \bibinfo {author} {\bibfnamefont {Y.}~\bibnamefont {Kanada-En’yo}}, \bibinfo {author} {\bibfnamefont {D.}~\bibnamefont {Lee}}, \ and\ \bibinfo {author} {\bibfnamefont {U.-G.}\ \bibnamefont {Meißner}},\ }\href {http://dx.doi.org/10.1103/RevModPhys.90.035004} {\bibfield  {journal} {\bibinfo  {journal} {Rev. Mod. Phys.}\ }\textbf {\bibinfo {volume} {90}},\ \bibinfo {pages} {035004} (\bibinfo {year} {2018})}\BibitemShut {NoStop}%
\bibitem [{\citenamefont {Bertulani}\ \emph {et~al.}(2002)\citenamefont {Bertulani}, \citenamefont {Hammer},\ and\ \citenamefont {van Kolck}}]{Bert2002}%
  \BibitemOpen
  \bibfield  {author} {\bibinfo {author} {\bibfnamefont {C.}~\bibnamefont {Bertulani}}, \bibinfo {author} {\bibfnamefont {H.-W.}\ \bibnamefont {Hammer}}, \ and\ \bibinfo {author} {\bibfnamefont {U.}~\bibnamefont {van Kolck}},\ }\href {\doibase 10.1016/s0375-9474(02)01270-8} {\bibfield  {journal} {\bibinfo  {journal} {Nucl. Phys. A}\ }\textbf {\bibinfo {volume} {712}},\ \bibinfo {pages} {37} (\bibinfo {year} {2002})}\BibitemShut {NoStop}%
\bibitem [{\citenamefont {Higa}\ \emph {et~al.}(2008)\citenamefont {Higa}, \citenamefont {Hammer},\ and\ \citenamefont {{van Kolck}}}]{HIGA08}%
  \BibitemOpen
  \bibfield  {author} {\bibinfo {author} {\bibfnamefont {R.}~\bibnamefont {Higa}}, \bibinfo {author} {\bibfnamefont {H.-W.}\ \bibnamefont {Hammer}}, \ and\ \bibinfo {author} {\bibfnamefont {U.}~\bibnamefont {{van Kolck}}},\ }\href {\doibase https://doi.org/10.1016/j.nuclphysa.2008.06.003} {\bibfield  {journal} {\bibinfo  {journal} {Nucl. Phys. A}\ }\textbf {\bibinfo {volume} {809}},\ \bibinfo {pages} {171} (\bibinfo {year} {2008})}\BibitemShut {NoStop}%
\bibitem [{\citenamefont {Hammer}\ \emph {et~al.}(2017)\citenamefont {Hammer}, \citenamefont {Ji},\ and\ \citenamefont {Phillips}}]{Chenhalo}%
  \BibitemOpen
  \bibfield  {author} {\bibinfo {author} {\bibfnamefont {H.-W.}\ \bibnamefont {Hammer}}, \bibinfo {author} {\bibfnamefont {C.}~\bibnamefont {Ji}}, \ and\ \bibinfo {author} {\bibfnamefont {D.~R.}\ \bibnamefont {Phillips}},\ }\href {\doibase 10.1088/1361-6471/aa83db} {\bibfield  {journal} {\bibinfo  {journal} {J. Phys. G: Nucl. Part. Phys.}\ }\textbf {\bibinfo {volume} {44}},\ \bibinfo {pages} {103002} (\bibinfo {year} {2017})}\BibitemShut {NoStop}%
\bibitem [{\citenamefont {Bedaque}\ \emph {et~al.}(2003)\citenamefont {Bedaque}, \citenamefont {Hammer},\ and\ \citenamefont {van Kolck}}]{Bed2003}%
  \BibitemOpen
  \bibfield  {author} {\bibinfo {author} {\bibfnamefont {P.}~\bibnamefont {Bedaque}}, \bibinfo {author} {\bibfnamefont {H.-W.}\ \bibnamefont {Hammer}}, \ and\ \bibinfo {author} {\bibfnamefont {U.}~\bibnamefont {van Kolck}},\ }\href {\doibase 10.1016/j.physletb.2003.07.049} {\bibfield  {journal} {\bibinfo  {journal} {Phys. Lett. B}\ }\textbf {\bibinfo {volume} {569}},\ \bibinfo {pages} {159} (\bibinfo {year} {2003})}\BibitemShut {NoStop}%
\bibitem [{\citenamefont {van Kolck}(1998)}]{van_Kolck_1998}%
  \BibitemOpen
  \bibfield  {author} {\bibinfo {author} {\bibfnamefont {U.}~\bibnamefont {van Kolck}},\ }in\ \href {\doibase 10.1007/bfb0104898} {\emph {\bibinfo {booktitle} {Chiral Dynamics: Theory and Experiment}}}\ (\bibinfo  {publisher} {Springer Berlin Heidelberg},\ \bibinfo {year} {1998})\ p.~\bibinfo {pages} {62}\BibitemShut {NoStop}%
\bibitem [{\citenamefont {van Kolck}(1999)}]{van_Kolck_1999}%
  \BibitemOpen
  \bibfield  {author} {\bibinfo {author} {\bibfnamefont {U.}~\bibnamefont {van Kolck}},\ }\href {\doibase 10.1016/s0375-9474(98)00612-5} {\bibfield  {journal} {\bibinfo  {journal} {Nucl. Phys. A}\ }\textbf {\bibinfo {volume} {645}},\ \bibinfo {pages} {273} (\bibinfo {year} {1999})}\BibitemShut {NoStop}%
\bibitem [{\citenamefont {Kaplan}\ \emph {et~al.}(1998)\citenamefont {Kaplan}, \citenamefont {Savage},\ and\ \citenamefont {Wise}}]{Kaplan_1998}%
  \BibitemOpen
  \bibfield  {author} {\bibinfo {author} {\bibfnamefont {D.~B.}\ \bibnamefont {Kaplan}}, \bibinfo {author} {\bibfnamefont {M.~J.}\ \bibnamefont {Savage}}, \ and\ \bibinfo {author} {\bibfnamefont {M.~B.}\ \bibnamefont {Wise}},\ }\href {\doibase 10.1016/s0370-2693(98)00210-x} {\bibfield  {journal} {\bibinfo  {journal} {Phys. Lett. B}\ }\textbf {\bibinfo {volume} {424}},\ \bibinfo {pages} {390} (\bibinfo {year} {1998})}\BibitemShut {NoStop}%
\bibitem [{\citenamefont {{Suzuki}}(1974)}]{1974PThPh..52..890S}%
  \BibitemOpen
  \bibfield  {author} {\bibinfo {author} {\bibfnamefont {Y.}~\bibnamefont {{Suzuki}}},\ }\href {\doibase 10.1143/PTP.52.890} {\bibfield  {journal} {\bibinfo  {journal} {Progr. Theor. Phys.}\ }\textbf {\bibinfo {volume} {52}},\ \bibinfo {pages} {890} (\bibinfo {year} {1974})}\BibitemShut {NoStop}%
\bibitem [{\citenamefont {Ronen}\ \emph {et~al.}(2006)\citenamefont {Ronen}, \citenamefont {Barnea},\ and\ \citenamefont {Leidemann}}]{Ronen:2005xs}%
  \BibitemOpen
  \bibfield  {author} {\bibinfo {author} {\bibfnamefont {Y.}~\bibnamefont {Ronen}}, \bibinfo {author} {\bibfnamefont {N.}~\bibnamefont {Barnea}}, \ and\ \bibinfo {author} {\bibfnamefont {W.}~\bibnamefont {Leidemann}},\ }\href {\doibase 10.1007/s00601-005-0147-6} {\bibfield  {journal} {\bibinfo  {journal} {Few Body Syst.}\ }\textbf {\bibinfo {volume} {38}},\ \bibinfo {pages} {97} (\bibinfo {year} {2006})}\BibitemShut {NoStop}%
\bibitem [{\citenamefont {Recchia}(2004)}]{recchia}%
  \BibitemOpen
  \bibfield  {author} {\bibinfo {author} {\bibfnamefont {P.}~\bibnamefont {Recchia}},\ }\href@noop {} {\emph {\bibinfo {title} {Halo effective field theory for the interaction of two alpha particles}}}\ (\bibinfo  {publisher} {Master Thesis},\ \bibinfo {year} {2004})\BibitemShut {NoStop}%
\bibitem [{\citenamefont {Ali}\ and\ \citenamefont {Bodmer}(1966)}]{ALI196699}%
  \BibitemOpen
  \bibfield  {author} {\bibinfo {author} {\bibfnamefont {S.}~\bibnamefont {Ali}}\ and\ \bibinfo {author} {\bibfnamefont {A.}~\bibnamefont {Bodmer}},\ }\href {\doibase https://doi.org/10.1016/0029-5582(66)90829-7} {\bibfield  {journal} {\bibinfo  {journal} {Nucl. Phys.}\ }\textbf {\bibinfo {volume} {80}},\ \bibinfo {pages} {99} (\bibinfo {year} {1966})}\BibitemShut {NoStop}%
\bibitem [{\citenamefont {Abramowitz}\ and\ \citenamefont {Stegun}(1964)}]{abramowitz+stegun}%
  \BibitemOpen
  \bibfield  {author} {\bibinfo {author} {\bibfnamefont {M.}~\bibnamefont {Abramowitz}}\ and\ \bibinfo {author} {\bibfnamefont {I.~A.}\ \bibnamefont {Stegun}},\ }\href@noop {} {\emph {\bibinfo {title} {Handbook of Mathematical Functions with Formulas, Graphs, and Mathematical Tables}}},\ \bibinfo {edition} {ninth dover printing, tenth gpo printing}\ ed.\ (\bibinfo  {publisher} {Dover},\ \bibinfo {address} {New York},\ \bibinfo {year} {1964})\BibitemShut {NoStop}%
\bibitem [{\citenamefont {Kievsky}\ \emph {et~al.}(2008)\citenamefont {Kievsky}, \citenamefont {Rosati}, \citenamefont {Viviani}, \citenamefont {Marcucci},\ and\ \citenamefont {Girlanda}}]{Kievsky_2008}%
  \BibitemOpen
  \bibfield  {author} {\bibinfo {author} {\bibfnamefont {A.}~\bibnamefont {Kievsky}}, \bibinfo {author} {\bibfnamefont {S.}~\bibnamefont {Rosati}}, \bibinfo {author} {\bibfnamefont {M.}~\bibnamefont {Viviani}}, \bibinfo {author} {\bibfnamefont {L.~E.}\ \bibnamefont {Marcucci}}, \ and\ \bibinfo {author} {\bibfnamefont {L.}~\bibnamefont {Girlanda}},\ }\href {\doibase 10.1088/0954-3899/35/6/063101} {\bibfield  {journal} {\bibinfo  {journal} {J. Phys. G: Nucl. Part. Phys.}\ }\textbf {\bibinfo {volume} {35}},\ \bibinfo {pages} {063101} (\bibinfo {year} {2008})}\BibitemShut {NoStop}%
\bibitem [{\citenamefont {Raynal}\ and\ \citenamefont {Revai}(1970)}]{RR}%
  \BibitemOpen
  \bibfield  {author} {\bibinfo {author} {\bibfnamefont {J.}~\bibnamefont {Raynal}}\ and\ \bibinfo {author} {\bibfnamefont {J.}~\bibnamefont {Revai}},\ }\href {\doibase 10.1007/BF02756127} {\bibfield  {journal} {\bibinfo  {journal} {Nuovo Cim. 68A: 612-22(21 Aug 1970).}\ } (\bibinfo {year} {1970}),\ 10.1007/BF02756127}\BibitemShut {NoStop}%
\bibitem [{\citenamefont {Marcucci}\ \emph {et~al.}(2020)\citenamefont {Marcucci}, \citenamefont {Dohet-Eraly}, \citenamefont {Girlanda}, \citenamefont {Gnech}, \citenamefont {Kievsky},\ and\ \citenamefont {Viviani}}]{10.3389/fphy.2020.00069}%
  \BibitemOpen
  \bibfield  {author} {\bibinfo {author} {\bibfnamefont {L.~E.}\ \bibnamefont {Marcucci}}, \bibinfo {author} {\bibfnamefont {J.}~\bibnamefont {Dohet-Eraly}}, \bibinfo {author} {\bibfnamefont {L.}~\bibnamefont {Girlanda}}, \bibinfo {author} {\bibfnamefont {A.}~\bibnamefont {Gnech}}, \bibinfo {author} {\bibfnamefont {A.}~\bibnamefont {Kievsky}}, \ and\ \bibinfo {author} {\bibfnamefont {M.}~\bibnamefont {Viviani}},\ }\href {https://www.frontiersin.org/journals/physics/articles/10.3389/fphy.2020.00069} {\bibfield  {journal} {\bibinfo  {journal} {Front. in Phys.}\ }\textbf {\bibinfo {volume} {8}},\ \bibinfo {pages} {69} (\bibinfo {year} {2020})}\BibitemShut {NoStop}%
\bibitem [{\citenamefont {Afzal}\ \emph {et~al.}(1969)\citenamefont {Afzal}, \citenamefont {Ahmad},\ and\ \citenamefont {Ali}}]{expdata}%
  \BibitemOpen
  \bibfield  {author} {\bibinfo {author} {\bibfnamefont {S.~A.}\ \bibnamefont {Afzal}}, \bibinfo {author} {\bibfnamefont {A.~A.~Z.}\ \bibnamefont {Ahmad}}, \ and\ \bibinfo {author} {\bibfnamefont {S.}~\bibnamefont {Ali}},\ }\href {\doibase 10.1103/RevModPhys.41.247} {\bibfield  {journal} {\bibinfo  {journal} {Rev. Mod. Phys.}\ }\textbf {\bibinfo {volume} {41}},\ \bibinfo {pages} {247} (\bibinfo {year} {1969})}\BibitemShut {NoStop}%
\bibitem [{\citenamefont {Kievsky}(1997)}]{KIEVSKYKOHN}%
  \BibitemOpen
  \bibfield  {author} {\bibinfo {author} {\bibfnamefont {A.}~\bibnamefont {Kievsky}},\ }\href {\doibase https://doi.org/10.1016/S0375-9474(97)81832-5} {\bibfield  {journal} {\bibinfo  {journal} {Nucl. Phys. A}\ }\textbf {\bibinfo {volume} {624}},\ \bibinfo {pages} {125} (\bibinfo {year} {1997})}\BibitemShut {NoStop}%
\bibitem [{min()}]{minuit}%
  \BibitemOpen
  \href {\doibase http://cernlib.web.cern.ch/cernlib/} {\ http://cernlib.web.cern.ch/cernlib/}\BibitemShut {NoStop}%
\bibitem [{\citenamefont {Fedotov}\ \emph {et~al.}(2004)\citenamefont {Fedotov}, \citenamefont {Kartavtsev}, \citenamefont {Kochkin},\ and\ \citenamefont {Malykh}}]{PhysRevC.70.014006}%
  \BibitemOpen
  \bibfield  {author} {\bibinfo {author} {\bibfnamefont {S.~I.}\ \bibnamefont {Fedotov}}, \bibinfo {author} {\bibfnamefont {O.~I.}\ \bibnamefont {Kartavtsev}}, \bibinfo {author} {\bibfnamefont {V.~I.}\ \bibnamefont {Kochkin}}, \ and\ \bibinfo {author} {\bibfnamefont {A.~V.}\ \bibnamefont {Malykh}},\ }\href {\doibase 10.1103/PhysRevC.70.014006} {\bibfield  {journal} {\bibinfo  {journal} {Phys. Rev. C}\ }\textbf {\bibinfo {volume} {70}},\ \bibinfo {pages} {014006} (\bibinfo {year} {2004})}\BibitemShut {NoStop}%
\bibitem [{\citenamefont {Cullum}\ and\ \citenamefont {Willoughby}(2002)}]{doi:10.1137/1.9780898719192.ch2}%
  \BibitemOpen
  \bibfield  {author} {\bibinfo {author} {\bibfnamefont {J.~K.}\ \bibnamefont {Cullum}}\ and\ \bibinfo {author} {\bibfnamefont {R.~A.}\ \bibnamefont {Willoughby}},\ }\href {\doibase 10.1137/1.9780898719192} {\emph {\bibinfo {title} {Lanczos Algorithms for Large Symmetric Eigenvalue Computations}}}\ (\bibinfo  {publisher} {Society for Industrial and Applied Mathematics},\ \bibinfo {year} {2002})\BibitemShut {NoStop}%
\bibitem [{\citenamefont {Andreozzi}\ \emph {et~al.}(2003)\citenamefont {Andreozzi}, \citenamefont {Iudice},\ and\ \citenamefont {Porrino}}]{Andreozzi_2003}%
  \BibitemOpen
  \bibfield  {author} {\bibinfo {author} {\bibfnamefont {F.}~\bibnamefont {Andreozzi}}, \bibinfo {author} {\bibfnamefont {N.~L.}\ \bibnamefont {Iudice}}, \ and\ \bibinfo {author} {\bibfnamefont {A.}~\bibnamefont {Porrino}},\ }\href {\doibase 10.1088/0954-3899/29/10/302} {\bibfield  {journal} {\bibinfo  {journal} {J. of Phys. G: Nucl. Part. Phys.}\ }\textbf {\bibinfo {volume} {29}},\ \bibinfo {pages} {2319} (\bibinfo {year} {2003})}\BibitemShut {NoStop}%
\bibitem [{\citenamefont {Anderson}\ \emph {et~al.}(1999)\citenamefont {Anderson}, \citenamefont {Bai}, \citenamefont {Bischof}, \citenamefont {Blackford}, \citenamefont {Demmel}, \citenamefont {Dongarra}, \citenamefont {Du~Croz}, \citenamefont {Greenbaum}, \citenamefont {Hammarling}, \citenamefont {McKenney},\ and\ \citenamefont {Sorensen}}]{lapack99}%
  \BibitemOpen
  \bibfield  {author} {\bibinfo {author} {\bibfnamefont {E.}~\bibnamefont {Anderson}}, \bibinfo {author} {\bibfnamefont {Z.}~\bibnamefont {Bai}}, \bibinfo {author} {\bibfnamefont {C.}~\bibnamefont {Bischof}}, \bibinfo {author} {\bibfnamefont {S.}~\bibnamefont {Blackford}}, \bibinfo {author} {\bibfnamefont {J.}~\bibnamefont {Demmel}}, \bibinfo {author} {\bibfnamefont {J.}~\bibnamefont {Dongarra}}, \bibinfo {author} {\bibfnamefont {J.}~\bibnamefont {Du~Croz}}, \bibinfo {author} {\bibfnamefont {A.}~\bibnamefont {Greenbaum}}, \bibinfo {author} {\bibfnamefont {S.}~\bibnamefont {Hammarling}}, \bibinfo {author} {\bibfnamefont {A.}~\bibnamefont {McKenney}}, \ and\ \bibinfo {author} {\bibfnamefont {D.}~\bibnamefont {Sorensen}},\ }\href@noop {} {\emph {\bibinfo {title} {{LAPACK} Users' Guide}}},\ \bibinfo {edition} {3rd}\ ed.\ (\bibinfo  {publisher} {Society for Industrial and Applied Mathematics},\ \bibinfo {address} {Philadelphia, PA},\ \bibinfo {year} {1999})\BibitemShut {NoStop}%
\end{thebibliography}%
\end{document}